\documentclass[titlepage,11pt]{article}

\usepackage{longtable}

\usepackage[normalem]{ulem}
\usepackage[dvips]{graphicx}
\usepackage[myheadings]{fullpage}
\usepackage{pmetrika}
\usepackage{submit}
\usepackage{amssymb}
\usepackage{amsmath,nccmath}
\usepackage{babel}
\usepackage{adjustbox}
\usepackage{natbib}
\usepackage{import}
\usepackage{hyperref}

\newcommand{\Rset}{\mathbb{R}}
\newcommand{\limdistr}[1][n\to\infty]{\xrightarrow[n\rightarrow\infty]{\text{d}}}

\newcommand{\myexp}{\mathbb{E}} % expectation
\newcommand{\myvar}{\text{var}} % variance
\newcommand{\mycov}{\text{cov}} % covariance
\newcommand{\hp}{\mathcal{H}} % covariance

\newcommand{\fmat}[1][b]{F^{#1}}
\newcommand{\boot}[1]{\dot{#1}}

\newcommand{\Te}{S} % effective score

\newcommand{\Ts}{\tilde{S}} % standardized score

\newcommand{\multinormal}[1][n]{\mathcal{N}_{#1}}
\newcommand{\matrixnormal}[1][n]{\mathcal{MN}_{#1}}

\usepackage{xcolor}
\newcommand{\anna}[1]{\textcolor{black}{#1}}
\newcommand{\livio}[1]{\textcolor{black}{#1}}

\newcommand{\paolo}[1]{\textcolor{black}{#1}}

\begin{document}

\begin{titlepage}
	
	\title{Post-selection Inference in Multiverse Analysis (PIMA): an inferential framework based on the sign flipping score test
		%An inferential framework for Multiverse Analysis: the sign Flipping score test\\
		%OR\\
		%Valid $p$-hacking via flipscores test\\
		%OR\\
		%PIMA: Post-selection Inference in Multiverse Analysis
	}
	
	%%Disable \markright for your submission,
	%%if it includes any author names.
	%%Note: The "submit" package includes a running header.
	%\markright{\MakeLowercase{\textsc{}}}
	
	\author{Paolo Girardi$^1$, Anna Vesely$^2$, Dani\"el Lakens$^3$, Gianmarco Altoè$^4$, Massimiliano Pastore$^4$, Antonio Calcagnì$^4$, Livio Finos$^5$}
	
	\affil{$^1$Department of Environmental Sciences, Informatics and Statistics, Ca' Foscari University of Venice, Italy}
	\affil{$^2$Institute for Statistics, University of Bremen, Germany}
	\affil{$^3$Department of Industrial Engineering and Innovation Sciences, Eindhoven University of Technology, Netherlands}
	\affil{$^4$Department of Developmental Psychology and Socialisation, University of Padova, Italy}
	\affil{$^5$Department of Statistical Sciences, University of Padova, Italy}
	
	%\author{}
	%\affil{}
	
	\comment{\textbf{Contact Info}\\
		Paolo Girardi: paolo.girardi@unive.it; Anna Vesely: anna.vesely@unipd.it; Dani\"el Lakens: D.Lakens@tue.nl; Gianmarco Altoè: gianmarco.altoe@unipd.it; Massimiliano Pastore: massimiliano.pastore@unipd.it; Antonio Calcagnì: antonio.calcagni@unipd.it;
		Livio Finos: livio.finos@unipd.it.
	}
	
	\comment{\textbf{Founding}\\
		This research received no specific grant from any funding agency in the public, commercial, or not-for-profit sectors.}
	
	\comment{\textbf{Competing interests}\\
		The authors declare that they have no known competing financial interests or personal relationships that could have appeared to influence the work reported in this paper.}
	\newpage
	\comment{\textbf{Data Availability}\\
		All R code and data associated with the real data application are available at 
		\nolinkurl{https://osf.io/usrq7/?view_only=4c40978b0080496c98bb5b13592278b4}, 
		while further analyses can be developed trough the dedicated package Jointest \citep{Jointest}  available at \nolinkurl{https://github.com/livioivil/jointest}}
	%\thanks{I would like to thank .}
	\linespacing{1}
	\contact{Correspondence should be sent to
		\begin{flushleft}
			Paolo Girardi\break
			Address: Via Torino 155, 30172 Venezia-Mestre (VE), Italy\break
			E-Mail: paolo.girardi@unive.it  \break
		\end{flushleft} 
		\vspace{12cm}
		\vfill
	}
\end{titlepage}

\setcounter{page}{2}
\vspace*{2\baselineskip}

\RepeatTitle{Post-selection Inference in Multiverse Analysis (PIMA): an inferential framework based on the sign flipping score test}\vskip3pt

\linespacing{1.5}
%% ITEM 8 [See the "howto.tex" file.]
\abstracthead
\begin{abstract}
When analyzing data researchers make some decisions that are either arbitrary, based on subjective beliefs about the data generating process, or for which equally justifiable alternative choices could have been made. This wide range of data-analytic choices can be abused, and has been one of the underlying causes of the replication crisis in several fields. Recently, the introduction of multiverse analysis provides researchers with a method to evaluate the stability of the results across reasonable choices that could be made when analyzing data.  Multiverse analysis is confined to a descriptive role, lacking a proper and comprehensive inferential procedure. Recently, specification curve analysis adds an inferential procedure to multiverse analysis, but this approach is limited to simple cases related to the linear model, and only allows researchers to infer whether at least one specification rejects the null hypothesis, but not which specifications should be selected.
In this paper we present a Post-selection Inference approach to Multiverse Analysis (PIMA) which is a flexible and general inferential approach that accounts for all possible models, i.e., the multiverse of reasonable analyses. The approach allows for a wide range of data specifications (i.e. pre-processing) and any generalized linear model; it allows testing the null hypothesis of a given predictor not being associated with the outcome, by merging information from all reasonable models of multiverse analysis, and provides strong control of the family-wise error rate such that it allows researchers to claim that the null hypothesis can be rejected for each specification that shows a significant effect. The inferential proposal is based on a conditional resampling procedure. We formally prove that the type I error rate is controlled, and compute the statistical power of the test through a simulation study. Finally, we apply the PIMA  procedure to the analysis of a real dataset about coronavirus disease 2019 (COVID-19) vaccine hesitancy before and after the 2020 lockdown in Italy. We end with practical recommendations to consider when performing the proposed procedure.
\end{abstract}\vspace{\fill}\pagebreak

\section{Introduction}\label{sec:intro}

%\topic{p-hacking is a major cause of lack of replicability}

%\topic{Some important solutions have been proposed: Multiverse + Specification Curve + vibration of effects \cite{vibrationofeffects}}
% ----------------------------------------------------------------------------

%\topic{This is quite a result, but are informal, descriptive methods}

%\topic{A formally valid inferential framework is still missing. This paper aims to fill this theoretical gap.}

%\topic{In this paper we propose: 1) ensemble inference, 2) lower bound for True Discovery proportion (TDP), 3) model picking}

Real data analysis often provides many justifiable choices at each step of the analysis, such as how measurements are combined and transformed, how missing data and outliers are handled, and even the choice of a statistical model. Generally, there is not a single justifiable choice for each decision researchers need to make, and several justifiable options exist for each step of the data analysis \citep{gelman2014statistical}. As a consequence, raw data do not uniquely give rise to a single dataset for analysis. Instead, researchers are faced with a set of processed datasets, each of which is determined by a unique combination of choices -- a multiverse of datasets. As analyses performed on each dataset can lead to different results, the data multiverse directly implies a multiverse of statistical results.
In recent years, concerns have been raised about how researchers can abuse this flexibility in data analysis to increase the probability of observing a statistically significant result. \paolo{The reason researchers engage in such questionable research practices can be due both to the editorial's practice of predominantly publishing statistically significant results or the selection of findings that confirms the belief of the same authors \citep{begg1988publication,dwan2008systematic,fanelli2012negative}.} When researchers select and report the results of a subset of all possible analyses that produce significant results \citep{sterling1959publication,greenwald1975consequences, simmons2011false, brodeur2016star}, they dramatically increase the actual false-positive rates despite their nominal endorsement of a low type I error rate (e.g., $5\%$).
Two solutions have been proposed to deal with the problem of p-hacking. The first is to require researchers to specify their statistical analysis plan before they look at the raw data. Such preregistered studies control the type I error rate by reducing flexibility during the data analysis. Preregistration is easily implemented for replication studies, where researchers specify they will perform the same analysis as was performed in an earlier study. For more novel studies, preregistration can be difficult because researchers often lack sufficient knowledge to be able to foresee how they should deal with all possible decisions that need to be made when analyzing the data. The second solution acknowledges that it is often not feasible to specify a single analysis before the data has been collected, and instead promotes transparently reporting all possible analyses that can be performed. \cite{steegen2016increasing} introduced multiverse analysis which aims to use all reasonable options for data processing to construct a multiverse of datasets, and then separately perform the same analysis of interest on each of these datasets. The main tool used to interpret the output of a multiverse analysis is a histogram of p-values that summarizes all the p-values obtained for a given effect. Subsequently, researchers typically discuss the results in terms of the proportion of significant p-values. The procedure not only provides a detailed picture of the robustness or fragility of results across different choices for processing but also allows researchers to explore key choices that are most consequential in the fluctuation of their results.

Multiverse analysis represents an invaluable step toward a transparent science. The method has become increasingly popular since it was developed and has been applied in various experimental contexts, ranging from cognitive development and risk perception \citep{mirman2021advancing}, assessment of parental behavior \citep{modecki2020tuning}, and memory tasks \citep{wessel2020multiverse}. Although some of the applications remain confined to exploratory purposes with the scope to define brief guidelines for conducting a multiverse analysis \citep{dragicevic2019increasing,liu2020boba}, other studies aim to stimulate interest in this method as a robustness assessment for mediation analysis \citep{rijnhart2021assessing} or an exhaustive modeling approach \citep{frey2021identifying}. This research approach permits to exhibit the stability and robustness of discoveries, not just between different exclusion criteria or modifications of the variables, but between different decisions for all phases of the elaboration of the data. This feature can be particularly interesting and appealing from the perspective of the replicability crisis in quantitative psychology \citep{open2015estimating}, and as an attempt to increase the transparency and credibility of scientific results \citep{nosek2014method}. Multiverse analysis can therefore be extended not only to the pre-processing step but also to the methods used for the analysis (the “multiverse of methods”) \citep{harder2020multiverse}.

The explicit flexibility in multiverse analysis is not to be condemned as it reflects an effort to transparently describe the uncertainty about the best analysis strategy. However, if, on the one hand, the exploration of multiple analytic choices in data analysis must be advocated, on the other it is challenging to draw reliable inferences from such a large number of statistical analyses. Although most researchers have interpreted the results from multiverse analysis descriptively, while doing so it is extremely tempting to make claims about analyses that yield statistically significant results, and those that do not. However, a selective focus on a subset of statistically significant results once again introduces the problem of selective interference \citep{benjamini2020selective}, and can potentially inflate the rate at which claims about effects are false positives. 
Currently, the only method that allows researchers to make formal inferences in multiverse analysis is the specification curve analysis \citep{simonsohn2020specification}. Analogously to multiverse analysis, it requires researchers to consider the entire set of reasonable combinations of data-analytic decisions, called specifications; subsequently, these specifications are used jointly to derive a test for the null hypothesis of interest. If the null hypothesis is rejected, researchers can claim with a certain maximum error rate (e.g., $5\%$) that there exists at least one specification where the null hypothesis is false. In the most general case of non-experimental data, the inferential support is based on bootstrapping techniques and is valid only in linear regression models (LMs) without the possibility to provide a \anna{general} extension to other distributions for the dependent variable encompassed by generalized linear models (GLMs). More importantly, this methodology lacks a formal description of the statistical properties of the test, allows one to test only a single hypothesis, and the problem of controlling multiplicity when testing different hypotheses is not treated. A more formal study of the method's performance is provided in Sections \ref{sec:inference} and \ref{sec:sims}.

Because researchers are often interested in models more complex than LMs, want to explore several different processing steps, and possibly wish to investigate more null hypotheses together, it would be beneficial if more advanced analysis methods for multiverse analysis were developed. Such more advanced methods would, for instance, allow psychometricians to identify a group of predictors that are associated with a given outcome or neuroscientists to identify brain regions activated by a stimulus. To summarize, the multiverse analysis framework makes it possible to manage researchers degrees of freedom in the data analysis, but the literature still lacks a formal inferential approach that allows researchers to draw reliable inferences about (sets of) specific analyses included in multiverse analysis.

In this paper, we define the Post-selection Inference approach to Multiverse Analysis (PIMA) which is a flexible and general inference approach for multiverse analysis that accounts for all possible models, i.e., the multiverse of reasonable analyses. In the framework of GLMs, we consider the null hypothesis that a given predictor of interest is not associated with the outcome, i.e., that the corresponding coefficient is zero. Moreover, we assume researchers consider all reasonable models that are given by different choices of data processing. We provide a resampling-based procedure based on the sign-flip score test of \cite{hemerik2020robust} and \cite{stdScore} that allows researchers to test the null hypothesis by merging information from all reasonable models, and show that this framework permits inferences about the coefficient of interest on three different levels. First, considering the predictor of interest, we compute a {\it global p-value} considering all models, so researchers can state whether the coefficient is non-null in at least one of the models in multiverse analysis. Secondly, we compute individual {\it adjusted p-values} for each model and thus obtain the set of models where the coefficient is non-null. Since PIMA accounts for multiplicity, researchers are able to freely choose the preferred model post-hoc, after trying all models and seeing results. In other words, the procedure allows selective inference, but unlike p-hacking, researchers can select statistically significant analyses from the multiverse while controlling the type I error rate. Finally, we define a third inference strategy for multiverse analysis where researchers provide a lower confidence bound for the {\it true discovery proportion} (TDP), i.e., the proportion of models with a non-null coefficient. In this analysis, researchers cannot individually identify statistically significant models in the multiverse, but in some cases it may be more powerful to report the true discovery proportion than individual p-values. Finally, we argue that the method can be easily extended to the case of multiple hypotheses on different coefficients. The resulting procedure is general, flexible, and powerful, and can be applied to many different contexts. It is valid as long as all the considered models are reasonable and specified in advance, before carrying out the analysis.

The structure of the paper is the following. In Section \ref{sec:flipscore} we define the framework and construct the desired resampling-based test. Subsequently, in Section \ref{sec:inference} we use the test to make inference in the multiverse framework. Then we study the properties of the PIMA method and we apply it to real data in Sections \ref{sec:sims} and \ref{sec:dataanalysis}, respectively. We conclude with Section \ref{sec:conclusion} that contains a short remark of the main results, with some references to still open issues in multiverse analysis and practical recommendations to the PIMA methodology. All the analyses and simulations were implemented using the statistical software R \citep{R}.
All R code and data associated with the real data application are available at 
 \nolinkurl{https://osf.io/usrq7/?view_only=4c40978b0080496c98bb5b13592278b4}, 
while further analyses can be developed trough the dedicated package Jointest \citep{Jointest}  available at \nolinkurl{https://github.com/livioivil/jointest}.

% ----------------------------------------------------------------------------

\section{The  sign-flip score test}\label{sec:flipscore}
In the context of multiverse analysis, there is not a single pre-specified model, while we are interested in testing the effect of a given predictor on a response variable in the multiverse of possible models. In order to test the global null hypothesis that the predictor has no effect in any of the considered models, one needs to define a proper test statistic and its distribution under the null hypothesis. Finding a solution within the parametric framework represents a formidable challenge, due to the inherent dependence among the univariate test statistics, which in most of cases is very high and usually non-linear. The resampling-based approach usually provides a solution to this multivariate challenge. \anna{We will rely on the sign-flip score test of \cite{hemerik2020robust} and \cite{stdScore} to define an asymptotically exact test for the global null hypothesis of interest. In this section, we specify the structure of the models and introduce the sign-flip score test for a single model specification. In the next section, we will give a natural extension to the multivariate framework. Finally, we will show how to employ the procedure within the closed testing framework \citep{closed} to make additional inferences on the models.}
%In this section, we rely on the sign-flip score test of \cite{hemerik2020robust} and \cite{stdScore} to define an asymptotically exact test for the global null hypothesis of interest. First, we specify the structure of the models and introduce the sign-flip score test for a single model specification. Subsequently, we give a natural extension to the multivariate framework.

\subsection{Model specification}
%\anna{[At the moment the notation is: capital letters for matrices/vectors, and lowercase letters for scalars. The only exceptions are: 1) some numbers (e.g. the total number of permutations $B$) and 2) the test statistic $S$. Are we ok with that? Should I change it to make it more readable?\\
%Moreover, now we are not distinguishing from random variables and observed values, even though it should all be random with the exception of the test statistic that we use in Theorem 1. What we do is very similar to what Riccardo has in his arxiv.]}

We consider the framework of GLMs. Let $Y=(y_1,\ldots,y_n)^\top\in\Rset^n$ be $n$ independent observations of a variable of interest, which is assumed to belong to the exponential dispersion family with density of the form
\begin{equation*}
    \anna{h}(y_i, \theta_i, \phi_i)=
    \exp\left\{\frac{y_i\theta_i-b(\theta_i)}{a(\phi_i)}+c(y_i,\phi_i)\right\}\qquad (i=1,\ldots,n),
\end{equation*}
where $\theta_i$ and $\phi_i$ are the canonical and the dispersion parameter, respectively. According to the usual literature of GLMs \citep{agresti}, the mean and variance functions are
\begin{equation*}
\mu_i=E[y_i]=b'(\theta_i),\qquad v(\mu_i)=b''(\theta)=\frac{\myvar(y_i)}{a(\phi_i)}.
\end{equation*}

We suppose that the mean of $Y$ depends on an observed predictor of interest $X=(x_1,\ldots,x_n)^\top\in\Rset^n$ and $m$ other observed predictors $Z=(z_1,\ldots,z_n)^\top\in\Rset^{n\times m}$ through a non linear relation
\begin{equation*}
    g(\mu_i)=\eta_i=x_i\beta+z_i\gamma %\label{def:gmu}
\end{equation*}
where $g(\cdot)$ denotes the link function, $\beta\in\Rset$ is a parameter of interest, and $\gamma\in\Rset^m$ is a vector of nuisance parameters.

Finally we define the following \anna{$n\times n$ matrices} , that will be used in the next sections:
\begin{align*}
D&=\text{diag}\{d_i\}=\text{diag}\left\{\frac{\partial \mu_i}{\partial \eta_i}\right\}\\
V&=\text{diag}\{v_i\}=\text{diag}\{\myvar(y_i)\}\\
W&=D V^{-1}D.
\end{align*}

%From the log-likelihood function ${\ell}(\beta)$ we can derive the score function with respect to $\beta$ that is, in a compact matrix form,
%\begin{equation*}
%    {\ell}_*(\beta)=\frac{\partial {\ell}(\beta)}{\partial \beta}=X^TD{V}^{-1}(y-\mu),
%\end{equation*}
%where
%\begin{equation*}
%D=\text{diag}\left\{\frac{\partial \mu_i}{\partial \eta_i}\right\}=\text{diag}\{d_i\},\qquad V=\text{diag}\{\myvar(y_i)\}=\text{diag}\{v_i\}.
%\end{equation*}
%Furthermore, deriving the score function we can obtain the expected Fisher information
%\begin{equation*}
%    J(\beta)=\myexp\left(\frac{\partial^2 {\ell}(\beta)}{\partial \beta^2 }\right)=X^\top WX
%\end{equation*}
%where $W=D V^{-1}D$, and the expectation is with respect to the assumed model.

% ----------------------------------------------------------------------------

\subsection{Hypothesis testing for an individual model via sign-flip score test}
Given a model specified as in the previous section, we are interested in testing the null hypothesis $\hp:\beta=0$ that the predictor $X$ does not influence the response $Y$ with significance level $\alpha\in [0,1)$. Here $\gamma$ is estimated by $\hat{\gamma}$, and is therefore a vector of nuisance parameters. We consider the hypothesis $\beta=0$ for simplicity of exposition, however the sign-flip approach can be extended to the more general case $\beta=\beta_0$.

Relying on the work of \cite{hemerik2020robust}, \cite{stdScore} provide the sign-flip score test, a robust and asymptotically exact test for $\hp$ that uses $B$ random sign-flipping transformations. Even though larger values of $B$ tend to give more power, to have non-zero power it is sufficient to take $B\geq 1/\alpha$. Hence consider \anna{the $n\times n$ diagonal matrices $\fmat=\text{diag}\{f_i^b\}$, with $b=1,\ldots,B$.} The first is fixed as the identity $\fmat[1]=I$, and the diagonal elements of the others are independently and uniformly drawn from $\{-1,1\}$. Each matrix $\fmat$ defines a flipped effective score
\begin{equation}
    \Te^b=n^{-1/2}X^\top W^{1/2}(I-Q) V^{-1/2} \fmat (Y-\hat{\mu}) \label{eq:eff}
\end{equation}
where
\begin{equation*}
    Q=W^{1/2}Z(Z^\top WZ)^{-1}Z^\top W^{1/2}
\end{equation*}
is a particular hat matrix, symmetric and idempotent, and $\hat{\mu}$ is a $\sqrt{n}$-consistent estimate of the true value $\mu^*$ computed under $\hp$. \anna{In practical applications, if the matrices $D$ and $V$, and so $W$, are unknown, they can be substituted by $\sqrt{n}$-consistent estimates.}

This effective score may be written as a sum of individual contributions with flipped signs, as following.
\begin{equation}
    \anna{\Te^{b} =\frac{1}{\sqrt{n}} \sum_{i=1}^n f_i^b\,\nu_i,\qquad
    \nu_i = \left(x_i -  X^\top WZ (Z^\top WZ)^{-1}z_i\right)\,\frac{(y_i-\hat{\mu}_i)d_i}{v_i}}. \label{eq:sum_contr}
    %\frac{(y_i-\hat{\mu}_i)x_id_i}{v_i} - X^\top WZ (Z^\top WZ)^{-1} \frac{(y_i-\hat{\mu}_i)z_id_i}{v_i}.
\end{equation}
Here $\nu_i$ is the contribution  of the $i$-th observation to the effective score. \anna{The definition and properties of the contributions $\nu_i$ are explored in \citet{hemerik2020robust} and \citet{stdScore}, where they are denoted as $\nu_{\hat{\gamma},i}^*$ and $\tilde{\nu}_{i,\beta}^*$, respectively. }

An assumption is needed on the effective score computed when the true value $\gamma^*$ of the nuisance $\gamma$, and so the true value $\mu^*$ of $\mu$, are known. This quantity may be written analogously to \eqref{eq:eff} and \eqref{eq:sum_contr}, as
\begin{equation}
    \Te^{*b} = n^{-1/2}X^\top W^{1/2}(I-Q) V^{-1/2} \fmat(Y-\mu^*)=\frac{1}{\sqrt{n}}\sum_{i=1}^n f_i^b\,\nu^*_i. \label{eq:eff_star}
\end{equation}
\anna{In this case, the contributions $\nu_i^*$ are independent if $D$ and $V$ are known, and asymptotically independent otherwise \citep{hemerik2020robust}. The required assumption is a Lindeberg's condition that ensures that the contribution of each $\nu_i^*$ to the variance of $\Te^{b*}$ is arbitrarily small as $n$ grows. This can be formulated as following.}

\begin{assumption}\label{a:std}
As $n\rightarrow\infty$,
%\[\frac{1}{n}\sum_{i=1}^n\myexp(\nu_i^{*2})\longrightarrow c\]
\[\anna{\frac{1}{n}\sum_{i=1}^n\myvar(\nu_i^*)\longrightarrow c}\]
for some constant $c>0$. Moreover, for any $\varepsilon >0$
\[\frac{1}{n}\sum_{i=1}^n \myexp\left(\nu_i^{*2}\cdot\mathbf{1}\left\{\frac{|\nu_i^*|}{\sqrt{n}}>\varepsilon \right\} \right)\longrightarrow 0\]
where $\mathbf{1}\{\cdot\}$ denotes the indicator function.
\end{assumption}

Given this assumption, the sign-flip score test of \citet{stdScore} relies on the standardized flipped scores, obtained dividing each effective score \eqref{eq:eff} by its standard deviation:
\begin{equation}
    \Ts^b= \Te^b\, \myvar(\Te^b\,|\,\fmat)^{-1/2} \label{eq:std}
\end{equation}
where
\[\myvar(\Te^b\,|\,\fmat)=n^{-1} X^\top W^{1/2} (I-Q) \fmat (I-Q) \fmat (I-Q)W^{1/2}X + o_P(1).\]
\anna{The test is defined from the absolute values of the standardized scores, comparing the observed value $|\Ts^1|$ with a critical value obtained from permutations. The latter is $|\Ts|^{\lceil(1-\alpha)B\rceil}$, where $|\Ts|^{(1)}\leq\ldots\leq |\Ts|^{(B)}$ are all the sorted values and $\lceil\cdot\rceil$ denotes the ceiling function.}
%Let $|\Ts|^{(1)}\leq\ldots\leq |\Ts|^{(B)}$ be the sorted absolute values of the standardized scores, and let $\lceil\cdot\rceil$ denote the ceiling function. Then the test is defined as following.

\begin{theorem}[De Santis et al., 2022]\label{T:stdScore}
Under Assumption \ref{a:std}, the test that rejects $\hp$ when $|\Ts^1|>|\Ts|^{\lceil(1-\alpha)B\rceil}$ is an $\alpha$-level test, asymptotically as $n\rightarrow\infty$.
%\livio{under gaussian errors the (univariate) test is exact, for glm we have second-moment exactness. perhaps we should spend 2 words explaining what this mean (i.e. same as parametric score test)}
\end{theorem}

The test of Theorem \ref{T:stdScore} is exact in the particular case of LMs, and second-moment exact in GLMs. The second-moment exactness means that under $\hp$ the test statistics $\Ts^b$ do not necessarily have the same distribution, but share the same mean and variance, independently of the sign flip; this provides exact control of the type I error rate, for practical purposes, even for finite sample size. \anna{The only requirement is Assumption \ref{a:std}, that states that the variance of the score \eqref{eq:eff_star} is not dominated by any particular contribution.} Furthermore, the test is robust to some model misspecifications, as long as the mean $\mu$ and the link $g$ are correctly specified. In particular, under minimal assumptions, the test is still asymptotically exact for any generic misspecification of the variance $V$ \citep{stdScore}.

\subsection{\anna{Intuition behind the sign-flip score test}}
%\livio{ATTENZIONE: TUTTO DA RIVEDERE (TITOLO COMPRESO), paolo: ho fatto un primo editing} \anna{[ANNA: ho fatto qualche modifica, e soprattutto unificato la notazione sullo score e sulle hp nulle (che nel resto del testo sono $\hp$ e non $H_0$)]}

Although the formal definition of the sign-flip score approach may seem difficult to grasp, its meaning is quite intuitive. For the sake of clarity, we consider a simple example using a GLM with gaussian error and identity link, that can be easily reconducted to a multiple linear model. In this case, we have $W=D=I$ and $V=\sigma^2 I$, where $\sigma^2$ is the variance shared by every observation.
%\livio{From (\ref{eq:sum_contr}) and remembering that in gaussian homoscedastic linear model we have $W=D=I$ and  $V=\sigma^2 I$ (being $\sigma^2$ the variance shared by every observation):}
%\begin{equation}
%\begin{aligned}
%    \Te^{b} &= \sigma\sum_{i=1}^n \left(x_i -  z_i^\top (Z^\top Z)^{-1}X^\top\right) \left(y_i -  z_i^\top (Z^\top Z)^{-1}Y\right)\\ 
%    &=\sigma\sum_{i=1}^n\left(x_i -  z_i^\top  \hat{\beta}_x\right) \left(y_i -  z_i^\top \hat{\beta}_y\right)\\ 
%    &=\sigma\sum_{i=1}^n (x_i-\hat{x}_i)(y_i-\hat{y}_i).\label{eq:intuition}
%\end{aligned}
%\end{equation}
%\livio{(where we substitute $\hat{\mu}_i$ with $\hat{y}_i$ for mere coherence of notation with $\hat{x}_i=z_i^\top (Z^\top Z)^{-1}X^\top$)}
%\livio{forse i $ \hat{\beta}_x$ appesantiscono inutilmente la formula. ditemi anche se sia preferibile tenere $\hat \mu_i$ anzichè $\hat y_i$. (paolo: io terrei così) }
%\anna{[ANNA: credo (?) che ci sia un typo nella formula, dove bisogna dividere per $\sigma^2$ e non moltiplicare per $\sigma$. Inoltre manca la divisione per $\sqrt{n}$ e bisogna sistemare la distinzione tra permutati ($\Te^{b}$) e osservato ($\Te^{1}$). Io riscriverei il paragrafo più o meno così:]}\\
From \eqref{eq:sum_contr}, the observed and flipped effective scores can be written as
\begin{equation*}
    \Te^{1} =\frac{1}{\sqrt{n}}\sum_{i=1}^n \nu_i,\qquad \Te^{b} =\frac{1}{\sqrt{n}}\sum_{i=1}^n f_i^b\,\nu_i\qquad (b=2,\ldots,B)
\end{equation*}
where
\begin{equation}
\nu_i=\frac{1}{\sigma^2}(x_i-\hat{x}_i)(y_i-\hat{y}_i),\qquad\hat{x}_i=X^\top Z (Z^\top Z)^{-1}z_i,\qquad\hat{y}_i=\hat{\mu}_i=Y^\top Z (Z^\top Z)^{-1}z_i. \label{eq:intuition}
\end{equation}
In this perspective, the score can be interpreted as the sum of weighted residuals of $y_i -\hat{y}_i$, where the weights are the residuals $x_i-\hat{x}_i$.
A further interpretation is that the score is the sum of $n$ contributions, and these contributions are the residuals of $y_i$ predicted by $z_i$ multiplied by the residuals of $x_i$ predicted by $z_i$. In this sense, the score extends the covariance by moving from the empirical mean  (i.e., a model with the intercept only) to a full linear model.

To see things in practice, consider the following linear regression model 
%$$Y=\beta_0+\beta_1Z+\beta_2X + \varepsilon$$
%with $\varepsilon \sim \mathcal{N}(0,1)$,
\[Y=1+\beta X + \gamma Z+ \varepsilon,\qquad \varepsilon \sim \mathcal{N}_n(0,I)\]
and suppose we are interested in testing $\hp:\ \beta=0$. The predictors $X$ and $Z$ are generated from a multivariate normal with unit variance and covariance $0.80$.
We create two scenarios, sharing the same $X$ and $Z$, but with different response variable $Y$: the first scenario is generated under the null hypothesis $\hp$ ($\beta=0$, $\gamma=1$), while the second is generated under the alternative ($\beta=1$, $\gamma=1$). For each simulation we generate $n=100$ observations. We name the resulting datasets $H_0$ and $H_1$, respectively.

\begin{figure}
\centering
  \includegraphics[width=0.4\linewidth]{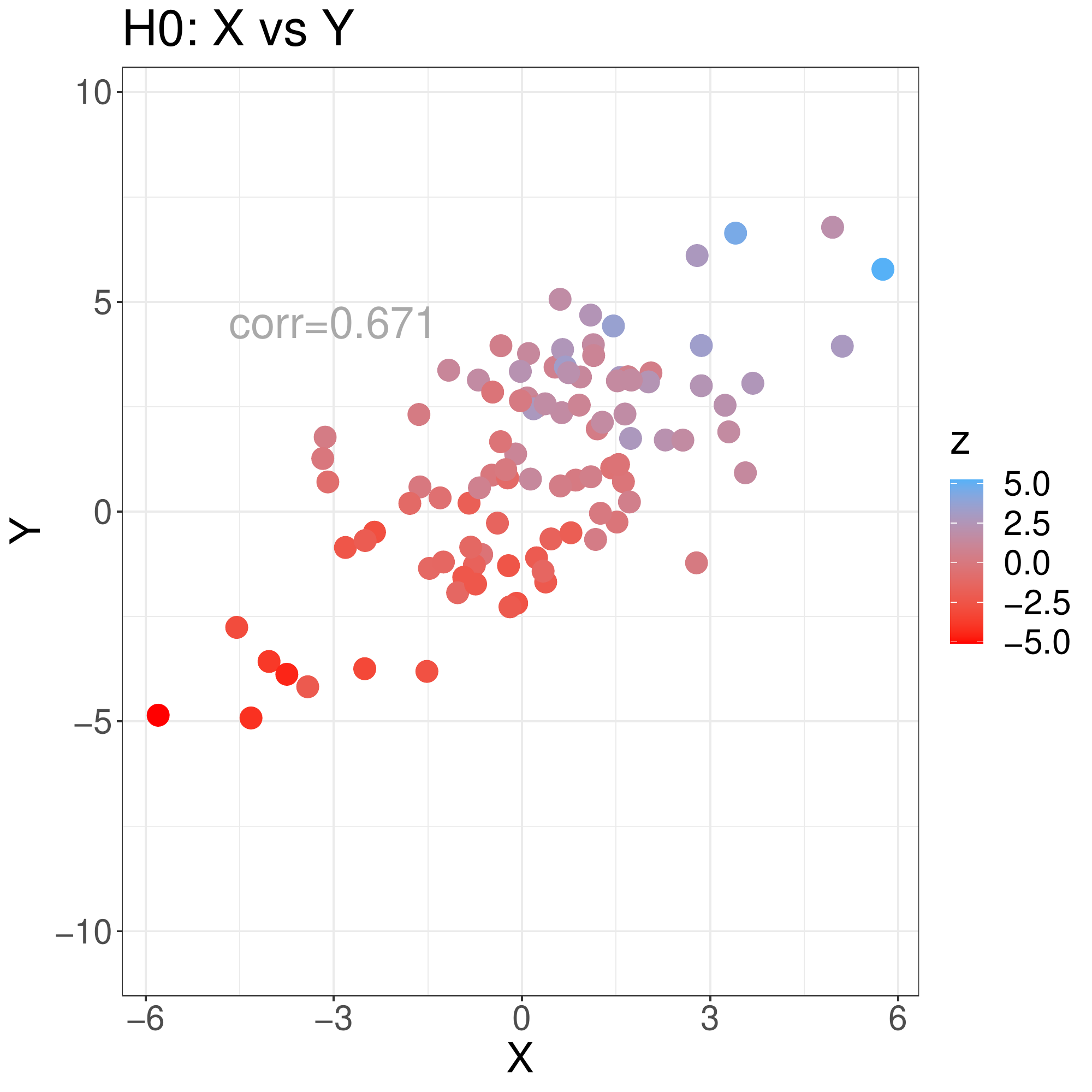}
    \includegraphics[width=0.4\linewidth]{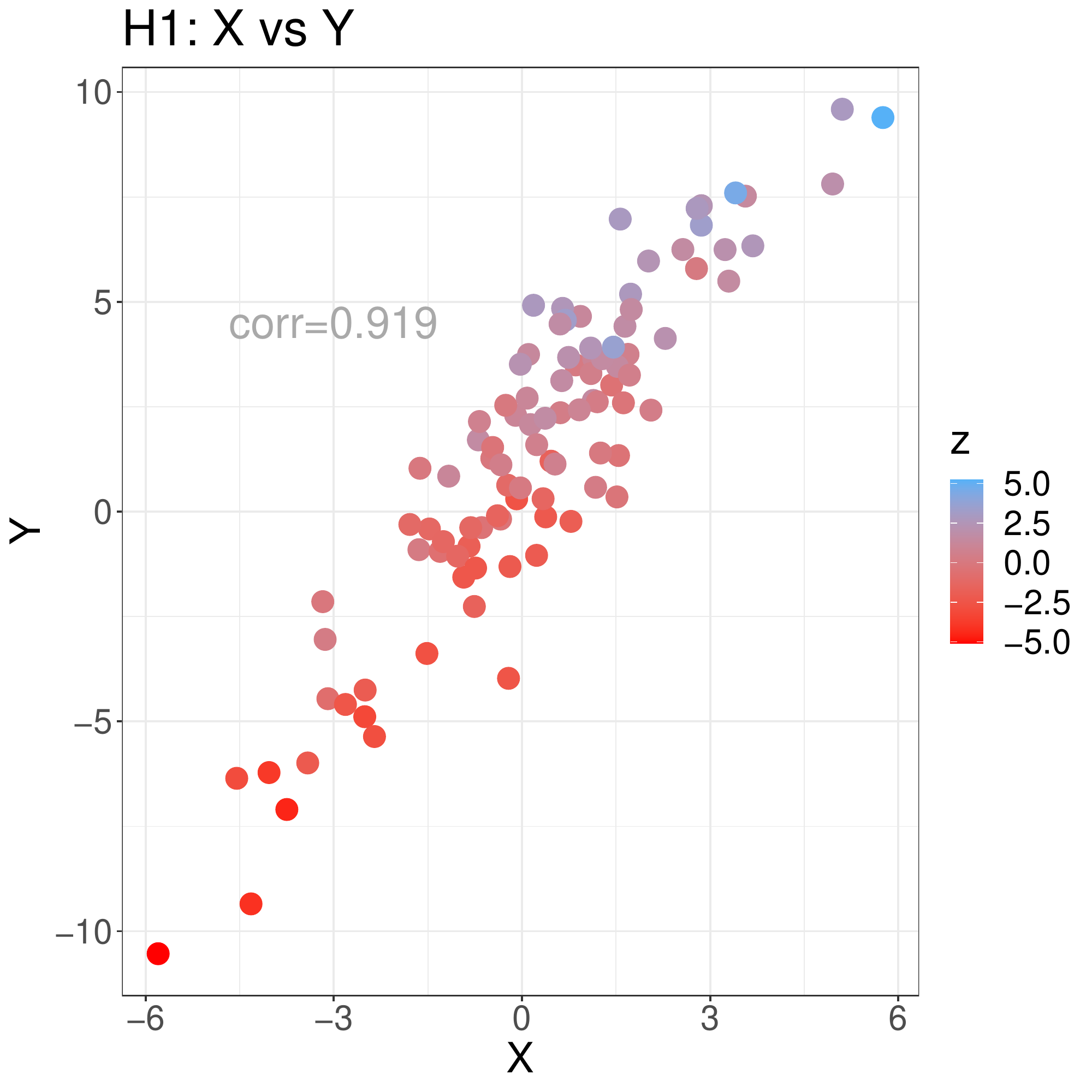}
  \caption{Simulated dataset under the scenario $H_0$ (left) and $H_1$ (right).}
  \label{fig:scatters_XYZ}
\end{figure}

\begin{figure}
\centering
    \includegraphics[width=0.4\linewidth]{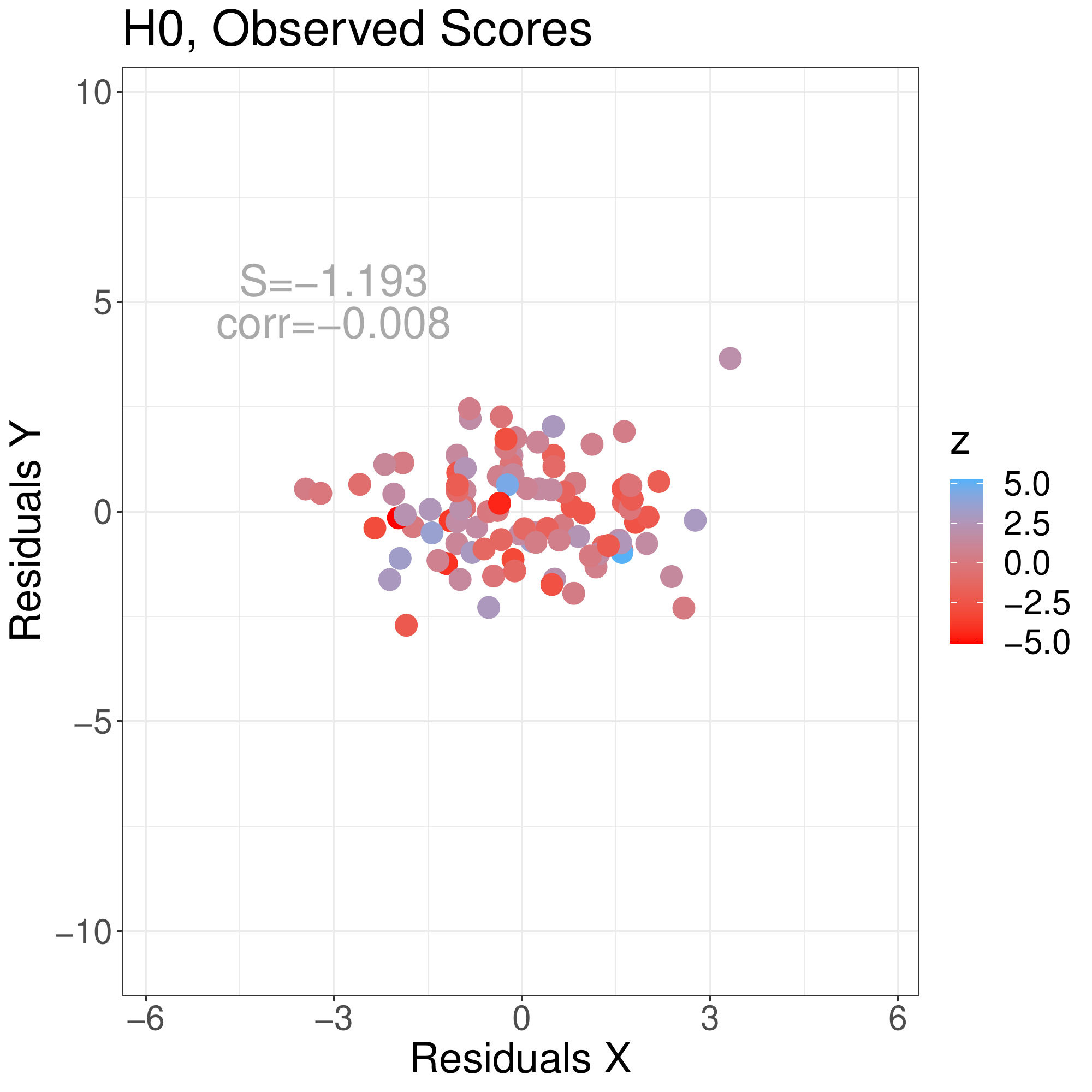}
    \includegraphics[width=0.4\linewidth]{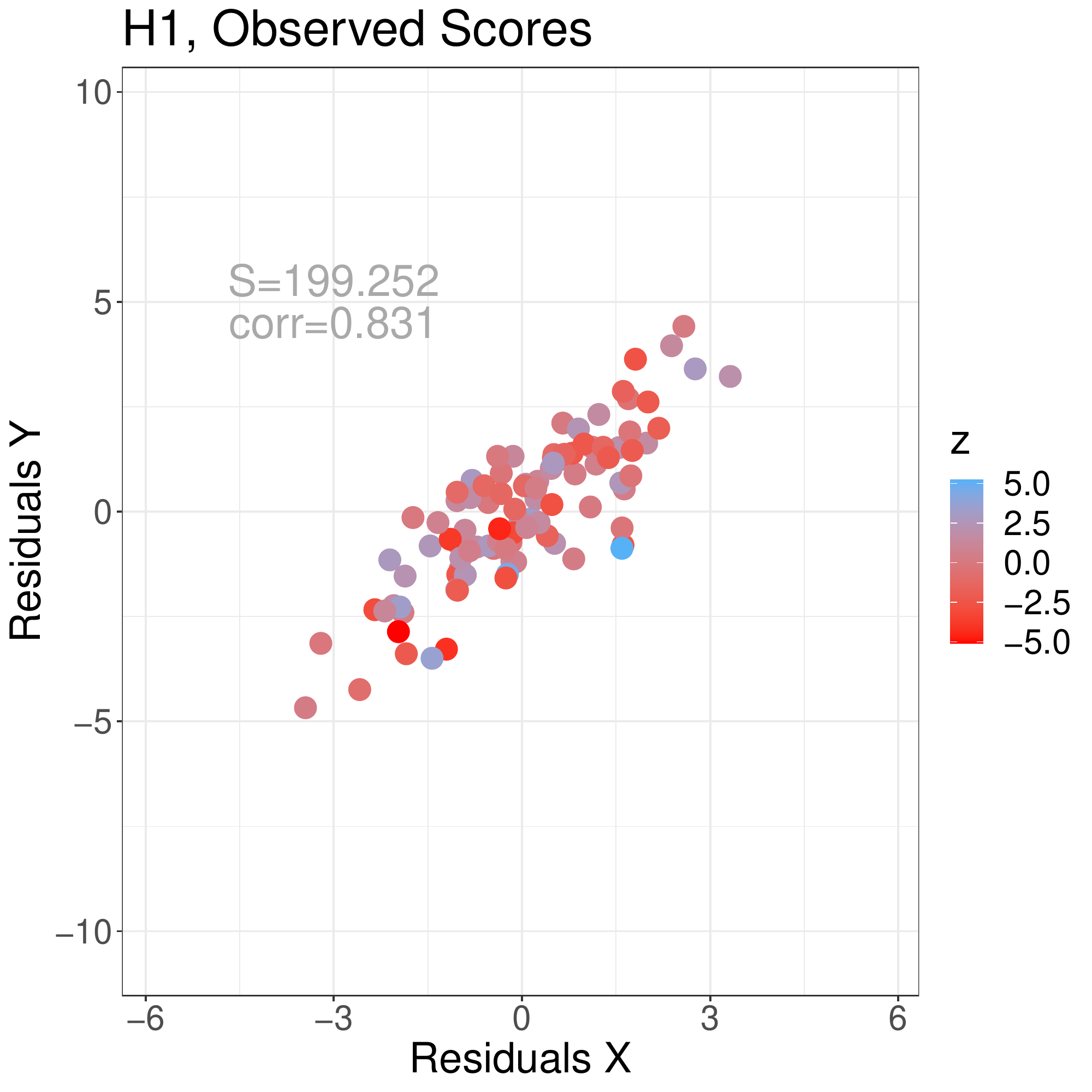}
    \includegraphics[width=0.4\linewidth]{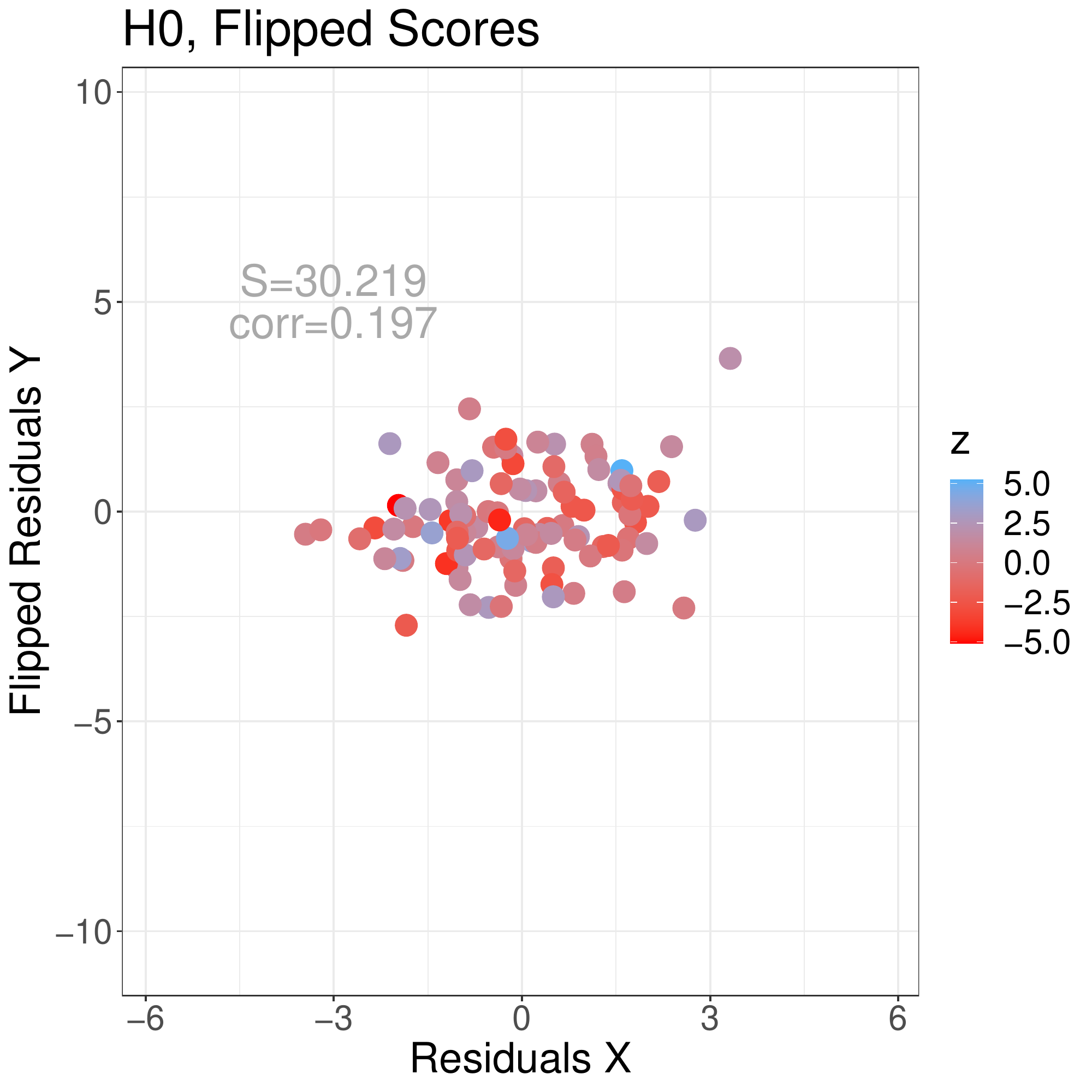}
    \includegraphics[width=0.4\linewidth]{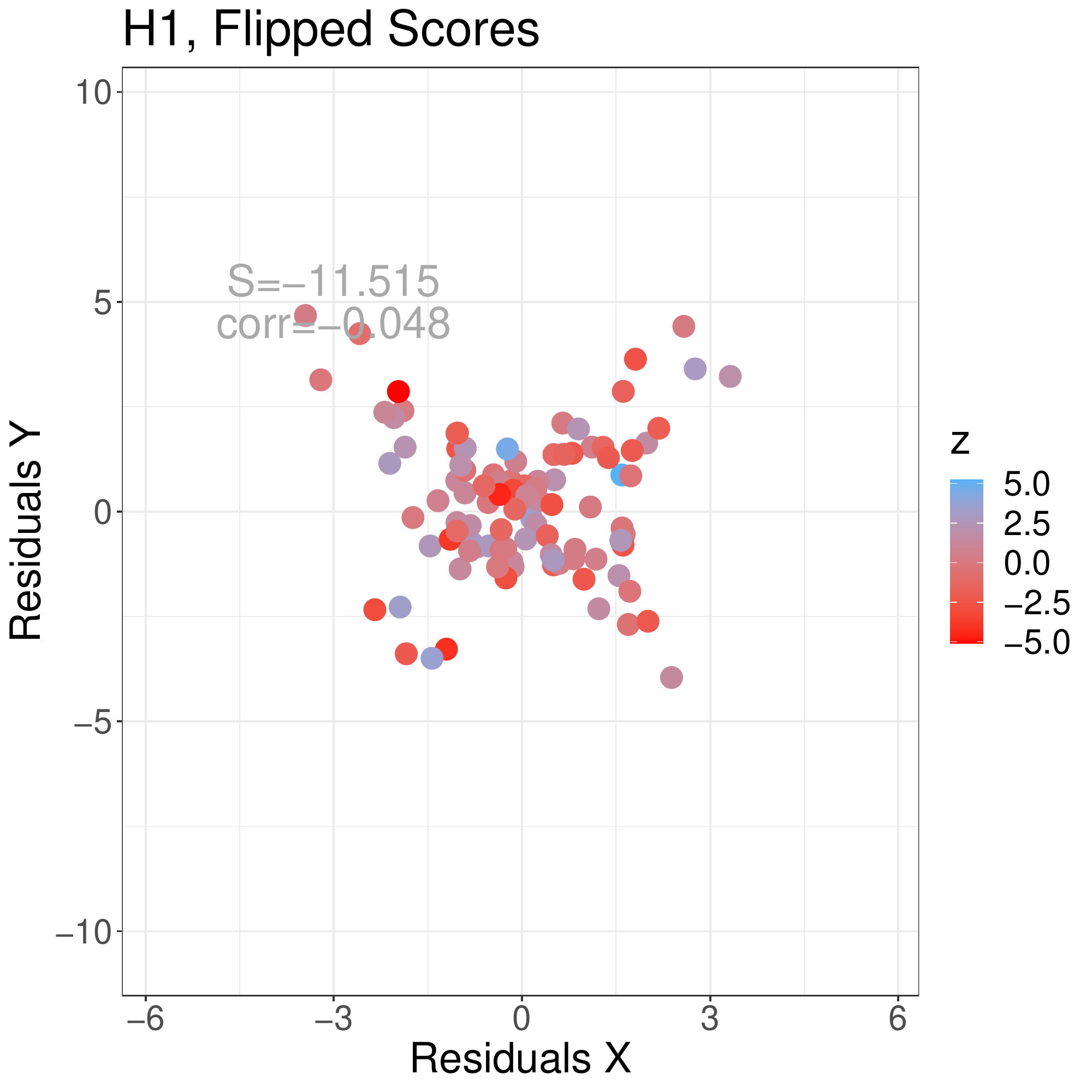}
  \caption{Observed (top) and flipped (bottom) distribution residuals of $Y$ vs. $X$ %under $H_0$ (left) and $H_1$ (right) scenario.
  in the datasets $H_0$ (left) and $H_1$ (right).}
  \label{fig:informal_flipscores}
\end{figure}

Examples of the scatter plots between $Y$ and $X$ considering also $Z$ by means of the color are given in Figure \ref{fig:scatters_XYZ}.
In both scenarios we see a positive correlation between $X$ and $Y$. From the color of the dots one can appreciate the positive dependence of $Z$ -- both -- with $X$ and $Y$; that is, more bluish dots corresponds to higher values of $Z$ and these appear where $X$ and $Y$ have higher values too (upper right corner).
Testing for the null hypothesis
%$H_0:\ \beta_1=0$
$\hp:\ \beta=0$, however, corresponds to testing the partial correlation between $X$ and $Y$, net to the effect of $Z$.
This partial correlation can be visually evaluated with a scatter plot of the residuals that form the $n$ addends of the observed score $\Te^1$ given in \eqref{eq:intuition}. These are shown in the two upper plots of Figure \ref{fig:informal_flipscores} for both the datasets $H_0$ (upper left) and $H_1$ (upper right). In these plots, the coordinates of each point are the values $x_i-\hat{x}_i$ and $y_i-\hat{y}_i$, and the observed score $\Te^1$ is obtained from the sum of the product of these coordinates $(x_i-\hat{x}_i)(y_i-\hat{y}_i)$.
After removing the effect of $Z$ from $X$ and $Y$, the scatter plot for $H_0$ does not show any relationship between the two variables, while this is still visible for $H_1$.
%From \eqref{eq:intuition}, the observed effective score $\Te^1$ is obtained from the sum of the product of the coordinates $(x_i-\hat{x}_i)(y_i-\hat{y}_i)$ of each point.}

The distribution of the effective score under the null hypothesis $\hp$ is obtained by computing a high number of flipped scores $\Te^b$. Each flipped score is determined by randomly flipping the signs of the score contributions $\nu_i$, and so of the residuals $(y_i-\hat{y}_i)$. The effect of these sign flips is visible in the scatter plots in the bottom of Figure \ref{fig:informal_flipscores}. The positive (partial) correlation of the $H_1$ dataset (top right) is destroyed by the random flips and is now approximately zero (bottom right). For the $H_0$ dataset a random flip maintains the observed correlation around zero. % The distribution under the null hypothesis $\hp$ is therefore obtained by repeatedly randomly flipping the residuals and computing the test statistic $\Te^b$.}

There are two further delicate details that may provide an additional value to the flip-scores approach: 1) the need for sign flips instead of permutations; 2) the need for the standardization step. One may notice that in the example proposed here, one may permute the residuals instead of flipping the signs. However, this is true only in the context of homoscedasticity, but would not be a valid option in the more general case of GLMs. Despite the intuition provided here fits to the GLM too, one has to bear in mind that the zero-centered contributions $\nu_i$ \eqref{eq:sum_contr} should be such that $\myvar(\nu_i)=\myvar(-\nu_i)$, while this would not hold when permuting the residuals $(y_i-\hat{y}_i)$.

%\livio{TODO: ANNA, facci il make up please!!}
The second relevant detail is the standardization step of \cite{stdScore}, introduced in \eqref{eq:std}. Due to the fact that the nuisance parameters $\gamma$ are unknown and must be estimated, the residuals $(y_i-\hat{y}_i)$ are independent only asymptotically. As a result, asymptotically the variances of the observed and permuted scores are equal, which ensures control of the type I error; however, this generally does not hold for finite sample sizes. %For this reason, one would not be allowed to flip the sign of the contributions independently, while doing so control of the type I error is ensured only asymptotically.For this reason, independently flipping the signs controls the type I error only asymptotically. 
The standardization step compensates for this different variability, guaranteeing exactness under the linear normal model and second-moment exactness in the more general GLM setting.
% ----------------------------------------------------------------------------

\section{PIMA: Post-selection Inference in the Multiverse Analysis}\label{sec:inference}

\subsection{Hypothesis testing in the multiverse via combination of sign-flip score tests} \label{sec:combine}

% METTERE NOTA SULLA NOTAZIONE TRA VETTORI, MATRICI E SCALARI:
% PER SEMPLIFICARE LA LETTURA NON FACCIAMO LA DISTINZIONE
% SI CAPISCE DAL CONTESTO
% cambiare B (piccolo) FORSE
% riguardare notazione: y_i piccoli se osservazioni, Y grande se vettore

In the previous section we presented an asymptotically exact test for a prefixed null hypothesis. Now we consider the framework of multiverse analysis, where we define $K$ plausible models, given by different processing of the data. Each model $k=1,\ldots,K$ can be characterized by different specifications of the response $Y_k$ (e.g., through outlier deletion or leverage point removal), of the predictors $X_k$ and $Z_k$ (e.g., combining and transforming variables), and of the link function $g_k$. Let $\beta_k$ be the coefficient of interest in model $k$, and define the null hypothesis $\hp_k:\beta_k=0$ analogously to the previous section. Then consider the global (i.e., multivariate) null hypothesis as the intersection of the $K$ individual hypotheses:
\begin{equation*}
    \hp=\bigcap_{k=1}^K \hp_k\,:\,\beta_k=0\text{ for all }k=1,\ldots,K.
\end{equation*}
This global hypothesis $\hp$ is true when the predictor of interest does not have any relation with the response within any of the $K$ models; it is false when such a relation exists in at least one of the models. To test $\hp$, we will extend the test of Theorem \ref{T:stdScore} similarly to the extension given in the case of the linear model of \cite{permMultisplit}.

To construct the desired global test, first we compute the flipped standardized scores \eqref{eq:std} for all models, using the same sign-flipping transformations. Hence we obtain $\Ts_1^b,\ldots,\Ts_K^b$ for $b=1,\ldots,B$. Intuitively, the $n$ scalar contributions $\nu_i$ in \eqref{eq:sum_contr} are now $n$ vectors of length $K$, each containing the contributions of the $i$-th observation to each one of the $K$ models. The same sign-flip for observation $i$, $f_i^b$, is therefore applied to the whole vector. This resampling strategy will ensure that the test has an asymptotically exact control of the type I error.

Subsequently we combine these flipped standardized scores through any function $\psi:\mathbb{R}^K\longrightarrow\mathbb{R}$ that is non-decreasing in each argument, such as the (weighted) mean and the maximum. This way we obtain the global test statistics
\begin{align}
T^b=\psi\left(|\Ts_1^b|,\ldots,|\Ts_K^b| \right)\qquad (b=1,\ldots,B). \label{def:combTs}
\end{align}
The following theorem gives a test for $\hp$ that relies on $T^1,\ldots,T^B$.

\begin{theorem}\label{T:multi}
Suppose that Assumption \ref{a:std} holds for all the considered models. Then the test that rejects $\hp$ when $T^1>T^{\lceil(1-\alpha)B\rceil}$ is an $\alpha$-level test, asymptotically as $n\rightarrow\infty$.
\end{theorem}

\begin{proof}
Throughout the proof, we will denote the $k$-th model adding a pedix $k$ to the corresponding quantities that vary between models. First, for simplicity of notation we consider only specifications that maintain the sample size, while we do not consider outlier deletion or leverage point removal. This way, the response vector $Y$ is the same between models.

Fix any $k\in\{1,\ldots,K\}$, and suppose that $\hp_k\,:\,\beta_k=0$ is true, so that the coefficient of interest is null in the $k$-th model. The flipped effective scores \eqref{eq:eff} are
\[\Te_k^b=n^{-1/2}X_k^\top W_k^{1/2}(I-Q_k) V_k^{-1/2} \fmat[b](Y-\hat{\mu}_k)\qquad (b=1,\ldots,B)\]
where
\[Q_k=W_k^{1/2}Z_k(Z_k^\top W_k Z_k)^{-1}Z_k^\top W_k^{1/2}\]
and $\hat{\mu}_k$ is a $\sqrt{n}$-consistent estimate of the true value $\mu_k^*$ computed under $\hp_k$. Consider the flipped effective scores computed when the true value $\gamma_k^*$ of the nuisance $\gamma_k$, and so the true value $\mu_k^*$ of $\mu_k$, are known as in \eqref{eq:eff_star},
\[
\Te_k^{*b}=n^{-1/2}X_k^\top W_k^{1/2}(I-Q_k) V_k^{-1/2} \fmat[b](Y-\mu_k^*)\qquad (b=1,\ldots,B).
\]
\cite{hemerik2020robust} show that $\Te_k^b$ e $\Te_k^{*b}$ are asymptotically equivalent as $n\rightarrow\infty$ (see the proof of Theorem 2).

Subsequently, assume that the global null hypothesis $\hp$ is true. Hence all individual hypotheses $\hp_k$ are true, and $\beta_k$ is null in all considered models. Consider the $KB$-dimensional vectors of effective scores
\begin{align*}
S &=(\Te_1^1,\ldots,\Te_1^B,\ldots,\Te_K^1,\ldots,\Te_K^B)^\top\\
S^* &= (\Te_1^{*1},\ldots,\Te_1^{*B},\ldots,\Te_K^{*1},\ldots,\Te_K^{*B})^\top
\end{align*}
that are asymptotically equivalent. For any couple of models $k,j\in\{1,\ldots,K\}$ and any couple of transformations $b,c\in\{1,\ldots,B\}$, we have
\begin{align*}
\myexp(\Te_k^{*b})&=0\\
\mycov(\Te_k^{*b},\Te_j^{*c})&=n^{-1}X_k^\top W_k^{1/2}(I-Q_k) V_k^{-1/2}\myexp\left(\fmat[b](Y-\mu_k^*)
(Y-\mu_j^*)^\top \fmat[c]\right) V_j^{-1/2}(I-Q_j)W_j^{1/2}X_j \\
&=
\begin{cases}
\xi_{kj}\quad\text{if } b=c\\
0\quad\text{otherwise}
\end{cases}
\end{align*}
where
\[\xi_{kj}=n^{-1}X_k^\top W_k^{1/2}(I-Q_k) V_k^{-1/2}\text{diag}\left((Y-\mu_k^*)
(Y-\mu_j^*)^\top\right) V_j^{-1/2}(I-Q_j)W_j^{1/2}X_j.\]

Note that $S^*$ can be written as the sum of $n$ independent vectors. As Assumption \ref{a:std} holds for all models, by the multivariate Lindeberg-Feller central limit theorem \citep{vaart}
\[S,S^*\limdistr \multinormal[NB]\left(\mathbf{0},\Xi\otimes I\right)\]
where $\mathcal{N}$ denotes the multivariate normal distribution, $\otimes$ is the Kronecker product, and
\begin{align*}
I\in\Rset^{B\times B},\qquad \Xi=\left(\lim_{n\rightarrow\infty}\xi_{kj}\right)\in\Rset^{K\times K}.
\end{align*}
Equivalently, we can say that
\begin{align*}
\begin{pmatrix}
\Te_1^1 & \ldots & \Te_K^1\\
\vdots &  & \vdots\\
\Te_1^B & \ldots & \Te_K^B
\end{pmatrix}
\limdistr \matrixnormal[s\times B]\left(0,I,\Xi\right)
\end{align*}
where $\mathcal{MN}$ denotes the matrix normal distribution. Hence the $B$ vectors of effective scores $(\Te_1^1,\ldots,\Te_K^1),\ldots, (\Te_1^B,\ldots,\Te_K^B)$ converge to i.i.d.~random vectors.

For each $k$, the standardized scores $\Ts_k^b$ are obtained dividing the effective scores $\Te_k^b$ by their standard deviation $\myvar(\Te_k^b\,|\,\fmat)^{1/2}$, as in \eqref{eq:std}. \cite{stdScore} show that these standard deviations are asymptotically independent of $b$ (see the proof of Theorem 2). Therefore, the $B$ vectors of the absolute values of standardized scores $(|\Ts_1^1|,\ldots,|\Ts_K^1|),\ldots,$ $(|\Ts_1^B|,\ldots,|\Ts_K^B|)$ converge to i.i.d.~random vectors. As a result, the combinations of their elements $T^1,\ldots,T^B$ defined in \eqref{def:combTs} converge to i.i.d.~random variables. Moreover, for each variable $k$ high values of $|\Ts_k^1|$ correspond to evidence against $\hp_k$ and $\psi$ is non-decreasing in each argument, and so high values of $T^1$ correspond to evidence against $\hp$. From \cite{hemerik2020robust} (see Lemma 1),
\[\lim_{n\to\infty}P\left(T^1 > T^{(\lceil (1-\alpha)B\rceil)}\right) = \frac{\lfloor\alpha B\rfloor}{B}\leq\alpha.\]

Finally, consider the more general case where we also allow for specifications that change the sample size, and so the response vector $Y_k$ may vary between models and have different lengths. The proof is written analogously to the previous one, with a slight modification of the sign-flipping matrices within each model. In model $k$ we use $\fmat_k$, obtained from $\fmat$ by removing the diagonal elements corresponding to the removed observations.
\end{proof}

Theorem \ref{T:multi} gives an asymptotically exact test for the global null hypothesis $\hp$ that the coefficient of interest is null in all considered models.  A global p-value can be immediately obtained as
\[p= \frac{1}{B} \sum_{b=1}^B \mathbf{1} \{T^b\geq T^1\}\]
\citep{exact}. 

An important role is played by the choice of the function $\psi$ that combines the flipped standardized scores to define the global test statistic \eqref{def:combTs}. There is a plethora of possible choices, any of them having different power properties in different settings. The most intuitive choices are the mean
\begin{align}
T_{\text{mean}}^b=\frac{1}{K}\sum_{i=1}^K |\Ts_k^b|\qquad (b=1,\ldots,B)\label{def:mean}
\end{align}
and the maximum
\begin{align}
T_{\text{max}}^b=\max_k |\Ts_k^b|\qquad (b=1,\ldots,B) \label{def:max}
\end{align}
%\livio{\@ Anna: define here also the MaxT since we use it also as global test. in the next section plese recall that maxT allows for dramatic shortcut of the closed testing.}
but the definition of the test remains flexible and general, allowing for several combinations. Other possible global test statistics can be obtained transforming the standardized scores $\Ts_k^b$ in p-values $p_k^b$, and then considering p-value combinations. The p-values can be defined either through parametric inversion of the scores or using ranks; we suggest this second choice, where
\[p_k^b= \frac{1}{B} \sum_{c=1}^B \mathbf{1} \{|\Ts_k^c|\geq |\Ts_k^b|\}\qquad (k=1,\ldots,K;\;b=1,\ldots,B).\]
Subsequently, the p-values can be combined with different methods such as those described and compared in \cite{pesarin}. We mention especially \cite{fisher}
\begin{align}
T_{\text{Fisher}}^b=-2 \sum_{k=1}^K \log p_k^b \qquad(b=1,\ldots,B)\label{def:fisher}
\end{align}
and Liptak/Stouffer \citep{liptak}
\begin{align*}
T_{\text{Liptak}}^b=- \sum_{k=1}^K \zeta(p_k^b)\qquad (b=1,\ldots,B)%\label{def:liptak}
\end{align*}
where $\zeta(\cdot)$ denotes the quantile function of the standard normal distribution.

\subsection{Post-selection Inference}\label{sec:postinference}

In the previous section, we considered different plausible specifications of a GLM and defined the global null hypothesis $\hp$ that a predictor of interest does not influence the response in any of these models. We constructed a test that combines the models' standardized scores to test $\hp$ with level $\alpha$, and so ensures weak control of the FWER. Therefore, if $\hp$ is rejected, we can state with confidence $1-\alpha$ that there exists at least one model where the predictor of interest has an influence on the response variable. In this section we show that the global test statistics $T^b$ defined in \eqref{def:combTs} can be used to make additional inferences on the models in two ways. We rely on the closed testing framework \citep{closed}, which has been proven to be the optimal way to construct multiple testing procedures, as all FWER, TDP, and related methods are either equivalent to or can be improved by it \citep{only}. It is based on the principle of testing different subsets of hypotheses by means of a valid $\alpha$-level local test, which in this case is the test of Theorem \ref{T:multi}.

First, to obtain adjusted p-values for each individual model, we apply the maxT-method of \cite{westyoung}, which corresponds to using as global test statistic the maximum defined in \eqref{def:max}. This procedure allows for a dramatic shortcut of the closed testing framework, and is fast and feasible even for high values of $K$ and $B$. The resulting p-values are adjusted for multiplicity, and so ensure strong control of the FWER. Researchers can postpone the choice of the preferred model after seeing the data, while still obtaining valid p-values; used in this way, the method allows researchers to make selective inferences. Where selective inference is a cause of the replication crisis when error rates are not controlled \citep{benjamini2020selective}, the PIMA procedure provides strong FWER control, allowing researchers to select a model after analyzing a multiverse of models without inflating the risk of a false positive.

Secondly, we can construct a lower ($1-\alpha$)-confidence bound for the proportion of models where the coefficient is non-null (TDP), using the general framework of \cite{genovese2} and \cite{exploratory} or, when the combining function $\psi$ can be written as a sum, the shortcut of \cite{sumsome}. The method allows to compute a confidence bound for the TDP not only for the whole set of models, but also simultaneously over all possible subsets without any adjustment of the $\alpha$-level. Simultaneity ensures that the procedure is not compromised by selective model selection. In this framework we are not able to individually identify statistically significant models, but in some cases reporting the TDP may be more powerful than individual adjusted p-values.

To conclude, the PIMA approach allows to make selective inference on the parameter of interest in the multiverse of models, providing not only a global p-value but also individual adjusted p-values and lower confidence bounds for the TDP of subsets of models. The PIMA procedure is exact only asymptotically in the sample size $n$; despite this, we will show through simulations that it maintains good control of the type I error even for small values of $n$. Furthermore, as shown in the real data analysis of Section \ref{sec:dataanalysis}, the same inference framework can be trivially extended to the case where we are interested in testing multiple parameters, i.e, where $\beta$ is a vector. Analogously to the extension from a single model to the multiverse, it is sufficient to define global test statistics \eqref{def:combTs} for all the individual parameters of interest using the same random sign-flipping transformations.

\subsection{Comparing PIMA with other proposals}
In this section we discuss and evaluate possible competitors to the PIMA procedure to test the global null hypothesis $\hp$. A first naive approach would be to rely on a parametric method. However, after computing a test for each model, there is the need to combine the univariate tests into a multivariate one. Since these tests coming from different specifications are generally non-independent and their dependence is very hard to model formally, the safest option is to use a Bonferroni correction. This approach has the invaluable advantage of simplicity, but has very low power in practice. This is mainly due to the strong correlation among model estimates that usually arises when different specifications of the same model are tested.

As mentioned in Section \ref{sec:intro}, the specification curve analysis of \cite{simonsohn2020specification} presents a first attempt to cast the descriptive approach of the multiverse analysis into an inferential framework. Two approaches are proposed. The first one relies on a naive permutation of the tested predictor followed by a refit of the models; the subsequent combination of the test statistics of each model follows the same logic exposed in Section \ref{sec:combine}. This method is only valid when the predictors are orthogonal, \anna{a setting that is typically limited to fully balanced experimental designs. Hence the method is no longer valid neither in experimental designs with unbalanced levels nor in non-experimental designs.}
%Since orthogonal predictors are typically limited to \livio{fully balanced} experimental designs, this approach is not very appealing in non-experimental designs.\livio{dobbiamo stressare qui o nella risposta che lui parla di experimental data ma in realta' intende fully balanced design? nel senso che basta che i livelli siano sbilanciati e le correlazioni sono non nulle?} \livio{ forse questa che segue (che ho proposto io) la toglierei visto che non abbiamo fatto simulazioni e ora che ho capito meglio i metodi potrebbe essere meno vera: (inoltre togliere la nuova riga} \anna{Io metterei una frase qui, e per il bootstrap sintetizzerei quello che abbiamo messo nella risposta}
%Furthermore, the method lacks statistical power when there are strong effects of the confounders in $Z$.
The second approach can be used in the more general case of non-experimental settings. \anna{It is originally defined for LMs and relies on the bootstrap method of \citet{flachaire1999abetter}}. For each specification, the model with the observed data is fitted (i.e., $y_i= \beta x_i + \gamma z_i + \varepsilon_i $), yielding the estimates of the parameters $\beta$ and $\gamma$. Then a
%new dependent variable
\anna{null response} $\boot{y}_i$ is generated by subtracting the estimated effect of the predictor of interest $x_i$ on $y_i$: $\boot{y}_i = y_i – \hat{\beta} x_i = (\beta - \hat{\beta}) x_i + \gamma z_i + \varepsilon_i$, where $\hat{\beta}$ is the sample estimate of $\beta$. The random variable $\beta - \hat{\beta}$ has zero-mean, therefore a null distribution of $\hat \beta$ can be obtained by re-fitting the model on bootstrapped data $\boot{y}_i,\ x_i,\ z_i$.
%With $\boot{y}_i$ we re-fit the same model on bootstrap samples of $\boot{y}_i= \boot{\beta} x_i + \gamma z_i + \varepsilon_i $), then we generate the null distribution of $\boot{\beta}$ by boostrapping the data. 
The resulting bootstrapped distribution of $\hat{\beta}$ is used to compute the p-value for $\hp$. Subsequently, the same resampling scheme is applied to each specification, and the resulting p-values are merged through appropriate combinations: \anna{the median, Liptak/Stouffer \citep{liptak}, and the count of specifications that obtain a statistically significant effect.}
%such as Fisher \citep{fisher} and Liptak/Stouffer \citep{liptak}.
\anna{For the case of LMs, both the bootstrap method and PIMA are robust to heteroscedasticity \citep{flachaire1999abetter} and ensure asymptotic control of the type I error. However, while the univariate test of the bootstrap method is still only asymptotically exact, the sign-flip score test on which PIMA relies has exact univariate control. Finally, the bootstrap refits the model at each step, and so requires a substantially larger computational effort, as will be confirmed in simulations.}

\anna{The same bootstrap procedure of \citet{simonsohn2020specification} is then extended to the case of GLMs. However, this extension is not always valid in our view. For GLMs, the authors base the bootstrap on the definition of null responses of the form $\boot{y}_i=g^{-1}(g(y_i)-\hat \beta x_i)$, but this proposal results to be very problematic for some models. For instance, the same issue occurs considering the binomial logit-link model, for which $y_i\in\{0,1\}$ and $\boot{y}_i=\text{expit}(\text{logit}(y_i)-\hat \beta x_i)$. When the response is $y_i=0$, we have that $\text{logit}(0)=-\infty$ and so the null response is always $\boot{y}_i=0$, regardless of the value of $\hat \beta x_i$. Analogously, when $y_i=1$ we always obtain $\boot{y}_i=1$. This means that the effect of the tested variable is never removed when computing the null response, contrary to what happens in the case of the LM for which the method is originally defined. The same problem arises for other GLMs that lead to infinite values, such as the Poisson log-link model, for which $\boot{y}_i=\exp(\log(y_i)-\hat \beta x_i)$ is always 0 when $y_i=0$.}
%followed by a rounding step to get plausible values of the distribution (e.g., 0 or 1 values of a binomial GLM, integers for a Poisson model, etc.).
%\livio{questa è da rivedere coerentemente con il commento fatto nella risposta. parliamone prima a voce (terrei la posizione che sì, possiamo usarlo in alcuni glm tipo poisson, ma non in tutti (es binomial).}
%\paolo{In fact, this proposal results to be very problematic since in binomial GLM $p_i=P(y_i=1|x_i)$ and $g(p_i)=\log(\frac{p_i}{1-p_i})$ produces non-finite values, due to the rounding process to 1 or 0 of $\boot{p}_i$; this latter issue affects also the variance. On the other hand, in many cases avoiding rounding is not feasible since probability law for discrete random variables is not defined for decimal values.}
\anna{Thus, in the case of GLMs we discourage the use of the proposal of \cite{simonsohn2020specification}.}
\livio{The poor control of the type I error in practice will be shown in the simulation study of Section \ref{sec:sims}.}

\anna{Finally, specification curve analysis only explores weak control of the FWER, i.e., infers on the presence of at least one significant specification. We underline the importance of the post-selection inference step introduced in PIMA, which allows to determine which models have a significant effect and so to gain a better understanding on the entire analysis.}

\section{\anna{Simulations}}\label{sec:sims}
%\livio{
%TODO: prendiamo un modello molto simile a quello di URI, ma inseriamo una covariata nel sequente modo...\\
%We simulate under sample size $n\in\{100,250,500\}$. Where present, the covariate $Z$ has correlation with the latent variable $\rho_{Z X_{\ell}}=0.6$ and has coefficient $\gamma=2$. 
%\begin{itemize}
%\item Scenario 1: Linear Model with homoscedastic normal errors
%\item Scenario 2: binomial logit-link model
%\item Scenario 3: Poisson log-link model 
%\item Scenario 4: negative binomial log-link model (intercept=-2), while a Poisson log-ling model is fitted. This setting simulate a scenario where the variance is misspecified  (dispersion parameter $\theta$ is set equal to $\mu$ so that the variance $\mu+\mu^2/\theta=2\mu$ is twice the variance expected in a Poisson model) 
%\end{itemize}
%correlation with latent x and each observed x is 0.85\\
%5000 repliche, 250 bootstraps and flips\\
%TODO: perchè non prendiamo la sig count combination:\\
%Among the combination methods proposed by Simonshon et al we choose the Median (i.e. the median of the estimated coefficients in each resampling) and the Stouffer (i.e. the sum of z scores of the tests in  each resampling). We do not consider the Count.sig one since it implicitly involves one-sides alternatives. Dealing with. The control of the directional errors results to be a non trivial task in multiple testing \ref{shaffer1980control,finner1999stepwise} and deserves the effort of a more formal dissertation.
%}

The following simulation study aims to assess the control of type I error and to quantify the power of the global test of Theorem \ref{T:multi}, performing a comparison with the bootstrapped method of specification curve analysis \citep{simonsohn2020specification}. Adjusted p-values for individual specifications are not reported, since type I error control of the global test automatically ensures FWER control through the closed testing principle, as argued in Section \ref{sec:postinference}. 

We set a common framework for all simulations, building on the settings used for specification curve analysis. \citet{simonsohn2020specification} simulated data generating the response $Y$ from a latent variable $X^{\ell}$ through a GLM, then considered a multiverse analysis with five different models. Each model uses a different predictor $X_k$, which is taken as a proxy for the latent $X^{\ell}$ and is generated so that it is strongly correlated to it. We expand on this setting, adding a confounder $Z$ which is also correlated to $X^{\ell}$. The pipeline for the analysis is as following. First, we simulate the latent variable $X^{\ell}$ and the confounder $Z$ from a multivariate normal distribution with mean zero, unitary variance and covariance $\rho_{X^{\ell}Z}=0.6$. Then we generate the response $Y$ through a GLM, taking
\[g(\mu_i) = x^{\ell}_i\beta + z_i\gamma+ \gamma_0.\]
Finally, we consider a multiverse analysis with five models. For each model $k$, we generate a new predictor $X_k$ so that it has a correlation with the latent variable $\rho_{X^{\ell}X_k}=0.85$. Then we fit a GLM with $X_k$ as the predictor of interest and $Z$ as a confounder.

We consider four scenarios, where in the first three we fit the correct model, while in the last one the variance in the fitted model is misspecified as follows:
%\livio{TODO occhio qui ho cambiato}
\begin{enumerate}
\item LM with homoschedastic Gaussian errors: $ \gamma_0=0$, $ \gamma=2$, $\beta=0$ (under the null hypothesis) or $\beta=0.2$ (under the alternative hypothesis), homoscedastic normal errors with variance 1;
\item Binomial logit-link model: $ \gamma_0=0$, $ \gamma=2$, $\beta=0$ (under the null hypothesis) or $\beta=0.5$ (under the alternative hypothesis);
\item Poisson log-link model: $ \gamma_0=0$, $ \gamma=2$, $\beta=0$ (under the null hypothesis) or $\beta=0.08$ (under the alternative hypothesis);
\item Data are generated with a Negative Binomial log-link model: $ \gamma_0=-2$, $ \gamma=2$, $\beta=0$ (under the null hypothesis) or $\beta=0.25$ (under the alternative hypothesis), and dispersion parameter $\theta=\mu$ (so that the variance $\mu+\mu^2/\theta=2\mu$ is twice the variance expected in a Poisson model). In this case a Poisson log-link model is fitted. In this scenario, we assess the robustness of the methods to misspecification of the variance. %However, the link function is still correct, hence the estimates of the coefficients are unbiased, and the score under the null hypothesis $\mathcal{H}:\beta=0$ has mean zero.
\end{enumerate}

For each scenario, we apply different tests with the scope to assess both the type I error rate and the power, setting the coefficient of interest $\beta$ at 0 in the first case (null hypothesis $\hp$) and at non-null values in the second (alternative hypothesis). We start exploring the behavior of univariate tests, applying, for each of the five models, three different methods: the sign-flip score test of Theorem \ref{T:stdScore}, the bootstrapped method of the specification curve analysis \citep{simonsohn2020specification}, and a suitable parametric test (t-test for LM, Wald test for the other GLMs). Subsequently, we combine information from the five considered models. We apply the PIMA method, taking as global test statistic the mean \eqref{def:mean} and the maximum \eqref{def:max}. We also report results for the bootstrapped method \citep{simonsohn2020specification}, combining the individual specifications' p-values with Stouffer and the median. We do not consider the combination that counts the specifications having a statistically significant effect since it implicitly involves one-sided alternatives, and dealing with the control of directional errors is a non trivial task in multiple testing \citep{shaffer1980control,finner1999stepwise} which deserves the effort of a more formal dissertation. Finally, an additional parametric global test could be obtained from the Bonferroni combination of the five univariate parametric tests; however, this is not feasible in practice, as the Bonferroni method results to be extremely conservative.
%\livio{la verità è che usavo il codice di uri che faceva bonferroni sui suoi test e non su quelli parametrici. me ne sono accorto tardi per fare rigirare tutto. in caso possiamo far girare subito così quando ce lo chiedono li abbiamo già pronti.}

Throughout simulations, we vary the sample size $n\in\{100,250,500\}$. Furthermore, we use $B=250$ random sign-flipping transformations and bootstraps; the choice of using a relatively reduced number is motivated by the huge computational effort required by the bootstrap, which refits the model at each step. We remark that the number of the random resamplings -- bootstraps or sign-flips -- does not affect the control of the type I error \citep{exact,ramdas2023permutation}. Each scenario is simulated $5{,}000$ times. This implies a standard error around significance level $5\%$ equal to  $\sigma_{\text{err}}=\sqrt{0.05\cdot 0.95/5000}=0.003$, and so the limits in this case are $\alpha\pm 1.96\sigma_{\text{err}}=(0.044, 0.056)$.

\begin{figure}
  \includegraphics[width=\linewidth]{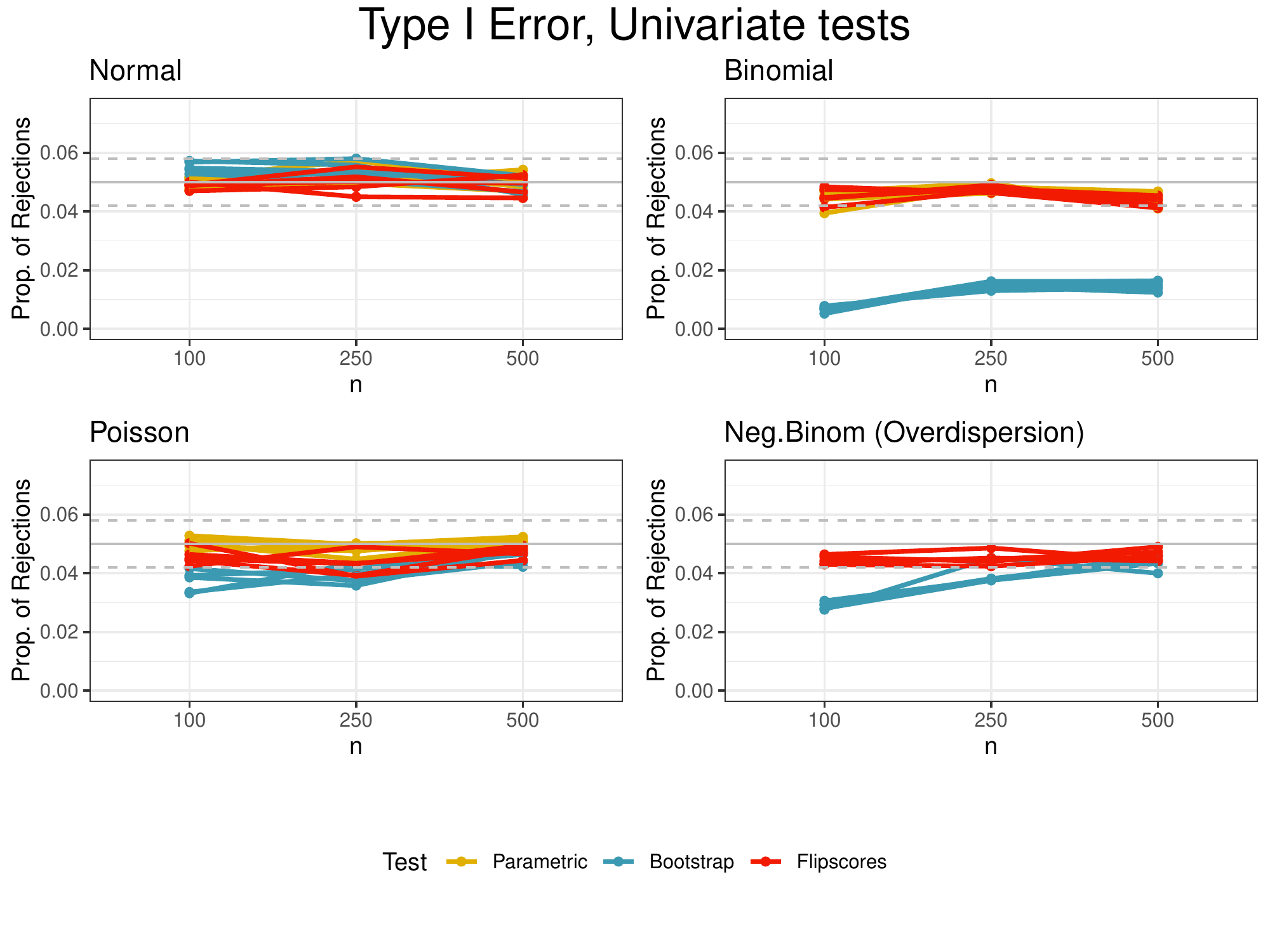}
  \caption{Simulations (univariate) under the null hypothesis $\hp:\beta=0$: empirical type I error of different methods for sample size $n\in\{100,250,500\}$ under four scenarios. For the Neg.~Binom scenario the empirical type I errors of the bootstrap approach exceed the upper limits of the ordinates -- ranging between  0.154 and 0.170 -- and are not shown. The dotted horizontal lines around $0.05$ correspond to the $95\%$ simulation error's limits.}
  \label{fig:simH0uni}
\end{figure}

\begin{figure}
  \includegraphics[width=\linewidth]{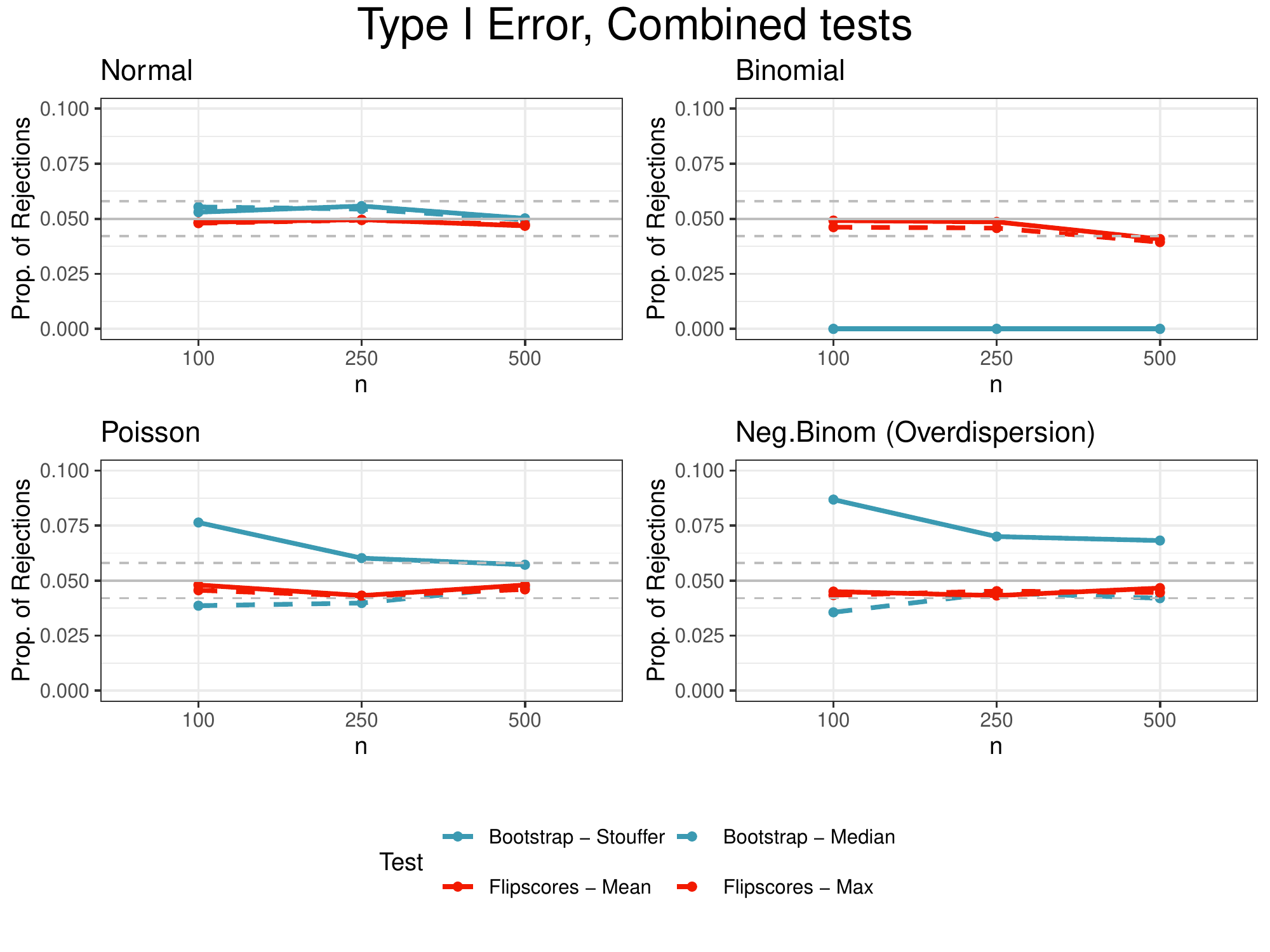}
  \caption{Simulations (combined) under the null hypothesis $\hp:\beta=0$: empirical type I error of different methods for sample size $n\in\{100,250,500\}$ under four scenarios.
  The dotted horizontal lines around $0.05$ correspond to the $95\%$ simulation error's limits.} 
  \label{fig:simH0combined}
\end{figure}

%\livio{TODO: in ultima passata per favore uniformate il nome flipscores (credo vada il plurale perchè flippiamo gli scores individuali, decidete se lo volete intero come il package R e quindi come nome proprio o flip scores separato o flip-scores non mi cambia molto.}

Figure \ref{fig:simH0uni} reports the type I empirical error rates for the different methods in the four scenarios. Each line reports the rejection proportion under the null hypothesis $\hp$ ($\beta=0$) for the five univariate models. Under the linear model (top-left plot), parametric tests and flipscores ones show a perfect control of the type I error as expected from the theory. Even the bootstrap method shows optimal behaviour, despite the control is ensured only asymptotically \citep{freedman1981bootstrapping}. In the Binomial (top-right) and Poisson (bottom-left) scenarios, the parametric and the flipscores are formally proven to have an asymptotic control of the type I error; the simulation confirms the good control in practice. 
The bootstrap approach shows a conservative behaviour in these settings, in particular with low sample sizes. Finally, in the Negative Binomial scenario, where a Poisson model with log link is fitted (bottom-right), the parametric model largely exceeds the putative $\alpha=0.05$ level, ranging between  0.154 and 0.170; this is not reported in the figure merely for graphical reasons. In the same setting, flipscores test performs well while the bootstrap remains conservative.

Figure  \ref{fig:simH0combined} shows the results of the multiverse of models under $\hp$. The boostrap and flipscores methods offer a comparable level of control for the linear model scenario. For GLMs, the boostrap appears to not adequately control for type I error, resulting too conservative in the Binomial scenario and exceeding the nominal level of $5\%$ in most cases for the Poisson and Negative Binomial scenarios.

%As a direct consequence of the lack of control of the type I error in the univariate parametric tests, the Bonferroni combination should not be considered for the heteroscedastic scenario. Restricting the evaluation to the homoscedastic scenario, the Bonferroni correction is very conservative due to its inability to account for the correlations among test statistics, which are typically very strong in a multiverse analysis.

\begin{figure}
  \includegraphics[width=\linewidth]{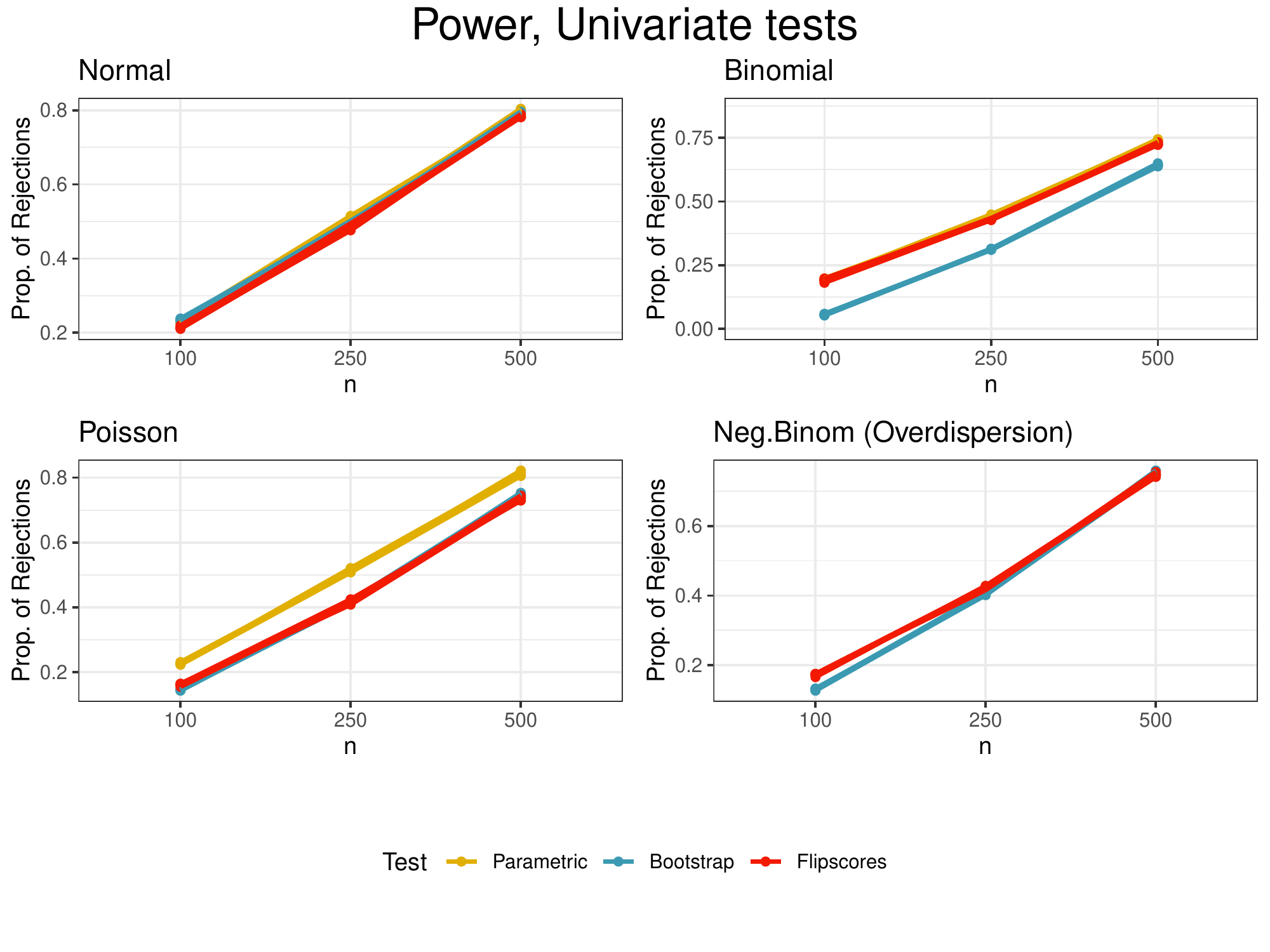}
  \caption{Simulations (univariate) under the alternative hypothesis ($\beta=1$): empirical power of the methods controlling for type I error for sample size $n\in\{100,250,500\}$ under four scenarios. The power of the methods that do not control the type I error -- i.e. exceeding the upper limit in the respective setting in Figure \ref{fig:simH0uni} --  are not shown.}
  \label{fig:simH1uni}
\end{figure}

%Table \ref{tab:sim_null} reports the empirical type I error for the true model (i.e. all linear predictors) and for different global tests in linear models (i.e. GLM with identity link and gaussian r.v.) with homoscedastic and heteroscedastic errors with increasing sample size $n$.
Considering only the methods controlling for the type I error in the previous univariate tests, the power increases as the sample size increases (Figure \ref{fig:simH1uni}); a slightly higher performance was observed for the parametric and flipscore procedures compared to the boostrap test in the Binomial scenario, and for the parametric method in the Poisson case. 
In the combined tests  the performances of the boostrap method (only with a Stouffer aggregation) are higher than the flipscores, while in the Binomial scenario the boostrap method fails. For the Poisson and the Negative Binomial scenario the simulations offer a similar power for both methods.

\begin{figure}
  \includegraphics[width=\linewidth]{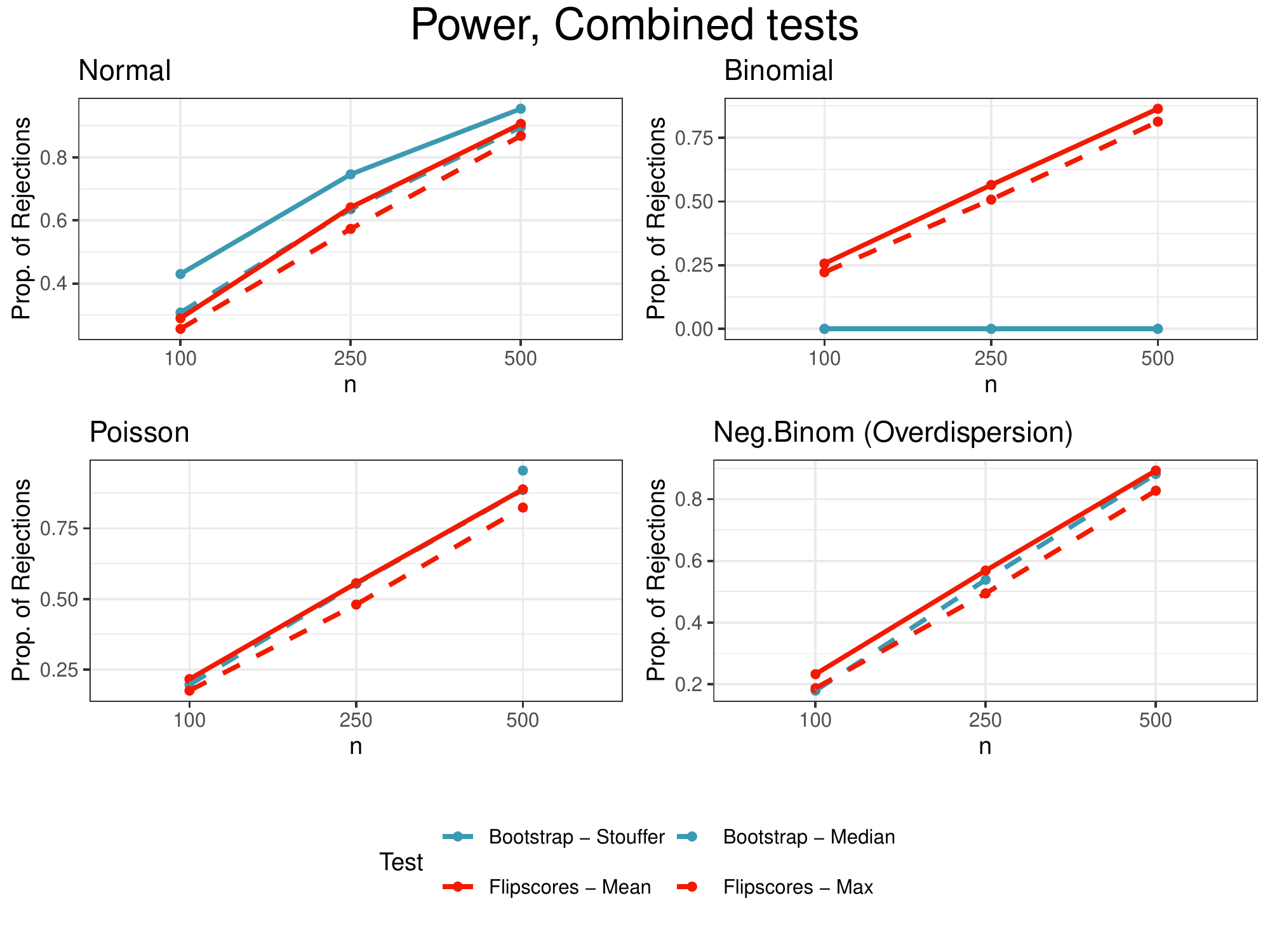}
  \caption{Simulations (combined) under the alternative hypothesis ($\beta=1$): empirical power of the methods controlling for type I error for sample size $n\in\{100,250,500\}$ under four scenarios. The power of the methods that do not control the type I error -- i.e. exceeding the upper limit in the respective setting in Figure \ref{fig:simH0combined} --  are not shown.}
  \label{fig:simH1combined}
\end{figure}

In conclusion, these simulations provide some insights to evaluate as the PIMA approach provides a general framework for the multiverse analysis, while the validity of the bootstrap method offers a limited superiority only under the scenario with linear model and Gaussian error. An exhaustive analysis in a wider range of settings would be of great interest, but would be very extensive. Indeed, the PIMA method is extremely general and flexible, as it applies to any GLM. A substantial number of scenarios could potentially be explored, considering different combinations of the characteristics studied here, as well as many others, such as the total number of predictors, their covariance, the nuisance parameters $\gamma$, the number and type of specifications, etc. In consequence, such an analysis is left for future work.

\section{Data analysis: The COVID-19 vaccine hesitancy dataset}\label{sec:dataanalysis}
% ----------------------------------------------------------------------------
\subsection{Description of the dataset}
The COVID-19 hesitancy dataset collected information on people's intention to get vaccinated and several characteristics \citep{caserotti2021}.  This survey was the first data collection that includes data on vaccine hesitancy before, during and after the lockdown in Italy that lasted from  March 8 until May 3, 2020.
%.  Previous analyses on this dataset evaluated how risk perception and some factors were associated with the decision to comply with COVID-19 vaccine acceptance in absence of a real vaccine availability \citep{caserotti2021, caserotti2022}. 
%In this analysis, we want to assess if during the lockdown people's hesitancy remained constant or varied since the different perceptions of the severity of COVID-19 respect the different phases of the epidemic outbreak.
The dataset is formed by a collection of voluntary respondents on the basis of a snow-ball sampling scheme. The willingness to be vaccinated was originally collected on a scale between 1 and 100; in this example, we mark as hesitant all people with an index below 100 ($n=1359$), while the others are marked as not hesitant ($n=909$). The main characteristics are reported in Table \ref{tab2}, in general and by the state of reluctance. Three variables were marginally associated with the status of hesitancy: \paolo{calendar} period, perceived risk of COVID-19 and doubts about vaccines. 
%In particular, the risk perception was one of the main factors that drove the hesitancy to get the vaccine.
%As in the previous analysis, we evaluate as three different specifications of this four confounding variables can affect the results on the period.

\begin{table}[htbp]
\begin{tabular}{lcccc}
\hline
\textbf{Characteristic} & \textbf{Overall} & \textbf{Hesitant} & \textbf{No hesitant} & \textbf{\textit{P}-value}\textsuperscript{2} \\ 
&  N = 2,268\textsuperscript{1}  &  N = 1,359\textsuperscript{1} & N = 909\textsuperscript{1} & \\
\hline
\textbf{Gender} &  &  &  & 0.065 \\
Female & 1,585 (70\%) & 930 (68\%) & 655 (72\%) &  \\
Male & 683 (30\%) & 429 (32\%) & 254 (28\%) &  \\
\textbf{Age} (years) & 35 (26, 49) & 35 (27, 48) & 35 (25, 51) & 0.4 \\ 
\textbf{Geographical area} &  &  &  & 0.2 \\
Center & 95 (4.2\%) & 65 (4.8\%) & 30 (3.3\%) &  \\
North & 2,015 (89\%) & 1,200 (88\%) & 815 (90\%) &  \\
South & 158 (7.0\%) & 94 (6.9\%) & 64 (7.0\%) &  \\
\textbf{Period} &  &  &  & $<0.001$ \\
Pre-lockdown & 845 (37\%) & 609 (45\%) & 236 (26\%) &  \\
Lockdown & 978 (43\%) & 494 (36\%) & 484 (53\%) &  \\
Post-lockdown & 445 (20\%) & 256 (19\%) & 189 (21\%) &  \\
\textbf{COVID-19 Perceived Risk}& 123 (80, 162) & 103 (62, 146) & 149 (110, 176) & $<0.001$ \\
\textbf{Vaccine doubts} & 8 (0, 25) & 11 (3, 40) & 2 (0, 10) & $<0.001$ \\ 
\textbf{Deprivation Index} & -0.69 (-1.61, 0.43) & -0.69 (-1.64, 0.43) & -0.69 (-1.49, 0.43) & 0.6 \\ 
\hline
\multicolumn{5}{l}{\textit{\textsuperscript{1}n (\%); Median (IQR)}}\\
\multicolumn{5}{l}{\textit{\textsuperscript{2}Pearson\textquotesingle{}s Chi-squared test; Wilcoxon rank sum test.}}\\
\hline
\end{tabular}
\caption{COVID-19 vaccine hesitancy: variables included in the analysis, overall and by hesitancy status.}
\label{tab2}
\end{table}

\subsection{Inferential approach}
We want to assess whether the doubts of the people before, during and after the Italian lockdown about a potential vaccine against COVID-19 remained constant or reported a substantial change, due to different perceptions of risk associated with COVID-19 contagion with respect to the different phases of the epidemic outbreak. To estimate the adjusted effect of the calendar period,  several confounders are taken into account: {\tt Covid\_perc\_risk}, COVID-19 Perceived risk, a scale defined combining different COVID-19 risk subscales (for further details, see \cite{caserotti2021}); {\tt doubts\_vaccine}, vaccine doubts on a 0-100 scale; {\tt age}, age in years; {\tt gender}, gender; {\tt age*gender}, interaction between age and gender; {\tt deprivation\_ index}, Italian Deprivation Index at the city of residence; {\tt geo\_are}, geographical area.

The variable under test is the date of the period of data collection {\tt Period} which has been recoded in a categorical variable with three levels according to the temporal window of the Italian lockdown: pre-lockdown ({\tt Pre}), during ({\tt Lockdown}) and post-lokdown ({\tt Post}). We are interested in all three possible comparisons, and their post-hoc corrected p-values. For each comparison, we fit a model with a zero-centered contrast that models the comparison of interest. As an example, to test the difference between {\tt Post} and {\tt Pre} we define $X$ as a variable having value 1 for {\tt Post}, -1 for {\tt Pre} and 0 for {\tt Lockdown}. The confounders $Z$ comprise a dummy variable for the level not involved in the comparison together with the above-mentioned confounders. %(Gender, Age, Geographical Area, COVID-19 Perceived risk, Vaccine doubts, and Deprivation Index.

%In particular, the risk perception was one of the main factors that drove the hesitancy to get the vaccine.
%As in the previous analysis, we evaluate as three different specifications of this four confounding variables can affect the results on the period.

%{\it \bf Basic model}
Having a dichotomous response $Y=\{ \textrm{not hesitant, hesitant}\}$, then recoded as $Y=\{1, 0\}$,
we use a GLM model with binomial response and logit link: 
\begin{align*}
y_i \sim Bernoulli(p_i),\qquad p_i \in (0,1)\\
g(p_i) = \log{\frac{p_i}{1-p_i}} = \alpha +\beta x_i +\gamma z_i.
\end{align*}

In order to implement a flexible approach, the relationship of the continuous predictors with the response is modeled also by basis of splines (B-splines). For each continuous predictors {\tt Covid\_perc\_risk}, {\tt doubts\_vaccine}, {\tt deprivation\_index} and {\tt age}, three transformations are tested: the natural variable, as well as a B-spline with three and four degree of freedom. Therefore, overall there are $K=3^4=81$ model specifications. For each comparison, e.g. {\tt Post} - {\tt Pre}, the $k$-th tested null hypothesis is defined in model $k$ as $\hp_k^{Post - Pre}:\ \beta_k^{Post - Pre}=0$. The global null hypothesis is the intersection of all null hypotheses, $\hp^{Post - Pre}=\cap_k^K \hp^{Post - Pre}_k$.

For each comparison we apply the PIMA framework with max-T combining function. Therefore we obtain: 1) a global p-value for the null hypothesis of no change over time (weak control of the FWER); 2) adjusted p-values for all individual models (strong control of the FWER);  3) lower confidence bounds for the TDP, i.e, the minimum proportion of models showing a significant difference. Furthermore, in this peculiar case we need to jointly test all possible pairwise comparisons: $\hp^{Post - Pre}\cap \hp^{Post - Lockdown} \cap \hp^{Lockdown - Pre}$. Accounting for the 81 model specifications, each one having three possible comparisons, we obtain 243 tests in total. The solution to this inferential problem is natural in the PIMA framework, as it is sufficient to define the closure set as the closure of the union of the univariate hypotheses of the three comparisons.

\subsubsection{Results}

We first report results for a parametric binomial model with linear predictors (i.e., natural variables, no B-spline used here) and two 0-centered contrasts variables that model the three-level {\tt Period} variable. Table \ref{tab:fitmod} reports the summary, while Table \ref{tab:posthocparam} shows the post-hoc Tukey correction for the three pairwise comparisons.  
% latex table generated in R 4.1.3 by xtable 1.8-4 package
% Fri Jul 01 20:30:33 2022
\begin{table}[h!]
\centering
\begin{tabular}{rrrrr}
  \hline
 & \textbf{Estimate} & \textbf{Std. Error} & \textbf{z value} & \textbf{Pr($>|z|$)} \\ 
  \hline
(Intercept) & -2.069 & 0.264 & -7.843 & 0.000 \\ 
  \paolo{Pre-lockdown} & -0.291 & 0.079 & -3.673 & 0.000 \\ 
  \paolo{Lockdown} & -0.089 & 0.072 & -1.236 & 0.216 \\ 
  Vaccine doubts & -0.036 & 0.003 & -13.029 & 0.000 \\ 
  Deprivation Index & -0.011 & 0.028 & -0.389 & 0.697 \\ 
  COVID-19 Perceived risk & 0.015 & 0.001 & 12.914 & 0.000 \\ 
  Age (+1 years) & 0.012 & 0.004 & 2.665 & 0.008 \\ 
  Gender [males] & 0.539 & 0.297 & 1.811 & 0.070 \\ 
  Geo. Area [North] & -0.026 & 0.176 & -0.149 & 0.881 \\ 
  Age*gender [males] & -0.016 & 0.007 & -2.106 & 0.035 \\ 
   \hline
\end{tabular}
\caption{\paolo{Summary of the estimated logistic regression model with logit link and linear confounders for COVID-19 vaccine hesitancy dataset. Period reference category is Post-lockdown}.}
\label{tab:fitmod}
\end{table}
% latex table generated in R 4.1.3 by xtable 1.8-4 package
% Fri Jul 01 20:29:20 2022
\begin{table}[h!]
\centering
\begin{tabular}{rrrrr}
  \hline
 & \textbf{Coefficients} & \textbf{Sigma} & \textbf{Tstat} & \textbf{$p$-values} \\ 
  \hline
Lockdown - Pre-lockdown & 0.202 & 0.124 & 1.634 & 0.230 \\ 
Post-lockdown - Lockdown & 0.469 & 0.138 & 3.392 & 0.002 \\ 
Post-lockdown - Pre-lockdown & 0.671 & 0.150 & 4.476 & 0.000 \\ 
   \hline
\end{tabular}
\caption{\paolo{Post-hoc pairwise comparisons with (Tukey) correction of the logistic regression model with logit link and linear confounders.}}
\label{tab:posthocparam}
\end{table}

% --------------------------------------------------
As introduced in the previous section, the multiverse analysis framework is built on the basis of three possible transformations for each continuous predictor ($81$ models) and also across the 3 comparisons ({\tt Pre} - {\tt Lockdown}, {\tt Pre} - {\tt Post}, {\tt Lockdown} -  {\tt Post}), leading to a multiverse of $81 \cdot 3 = 243$ models. Figure \ref{fig:scatter_raw} reports the results in a visual manner, while detailed results are reported in the Appendix. The common descriptive interpretation of a multiverse analysis allows to observe descriptively that the yellow and red clusters of tests yield p-values smaller than 0.05, but we cannot claim that these results are statistically significant due to the possibility that such a claim would have an unacceptably high false positive rate.

\begin{figure}
    \centering
    \includegraphics[scale=1]{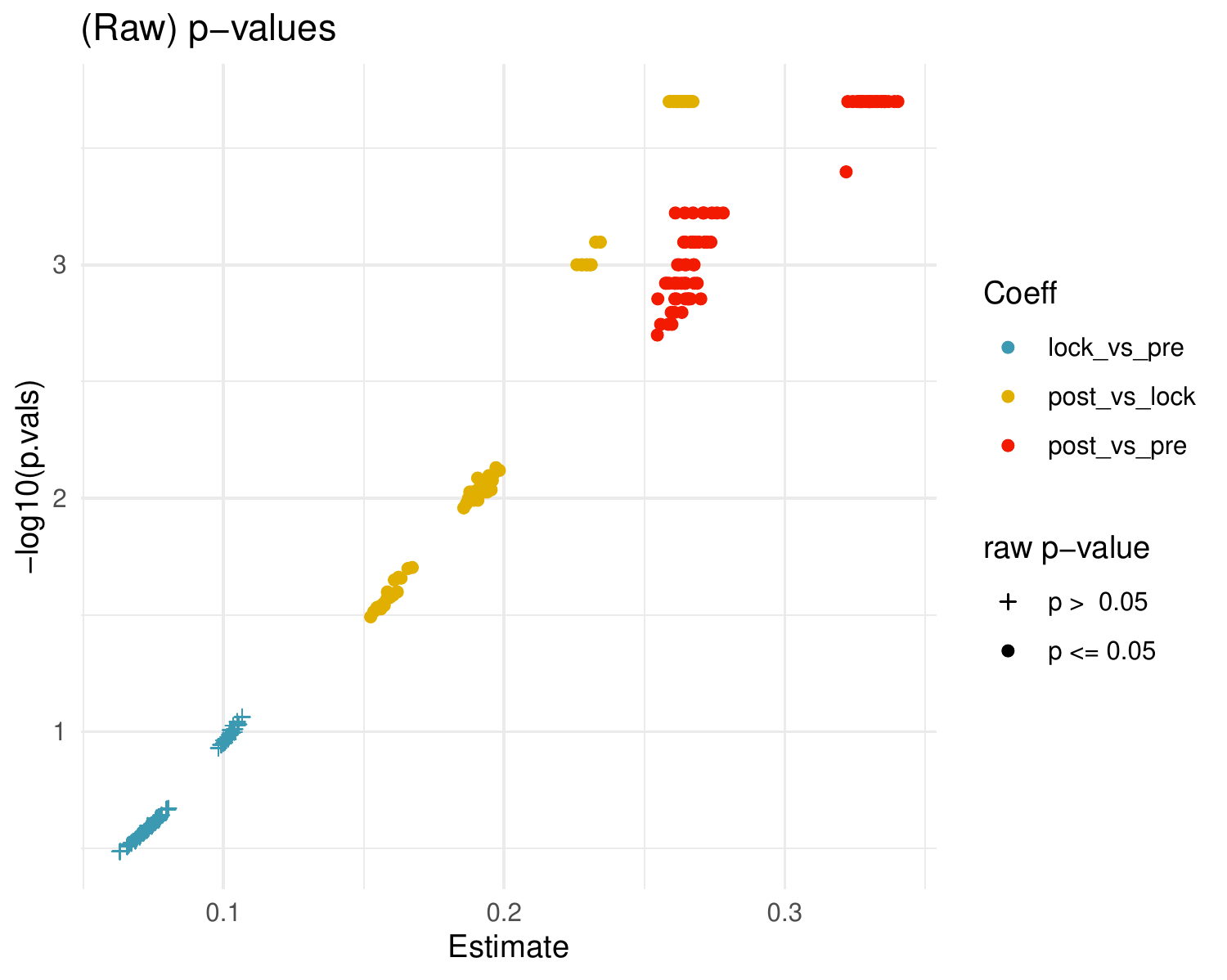}
\caption{
\paolo{Raw p-values vs. coefficient estimates of the post-hoc comparisons of 81 tested logistic models in the PIMA of COVID-19 vaccine hesitancy dataset.}}
    \label{fig:scatter_raw}
\end{figure}

\begin{figure}
    \centering
    \includegraphics[scale=1]{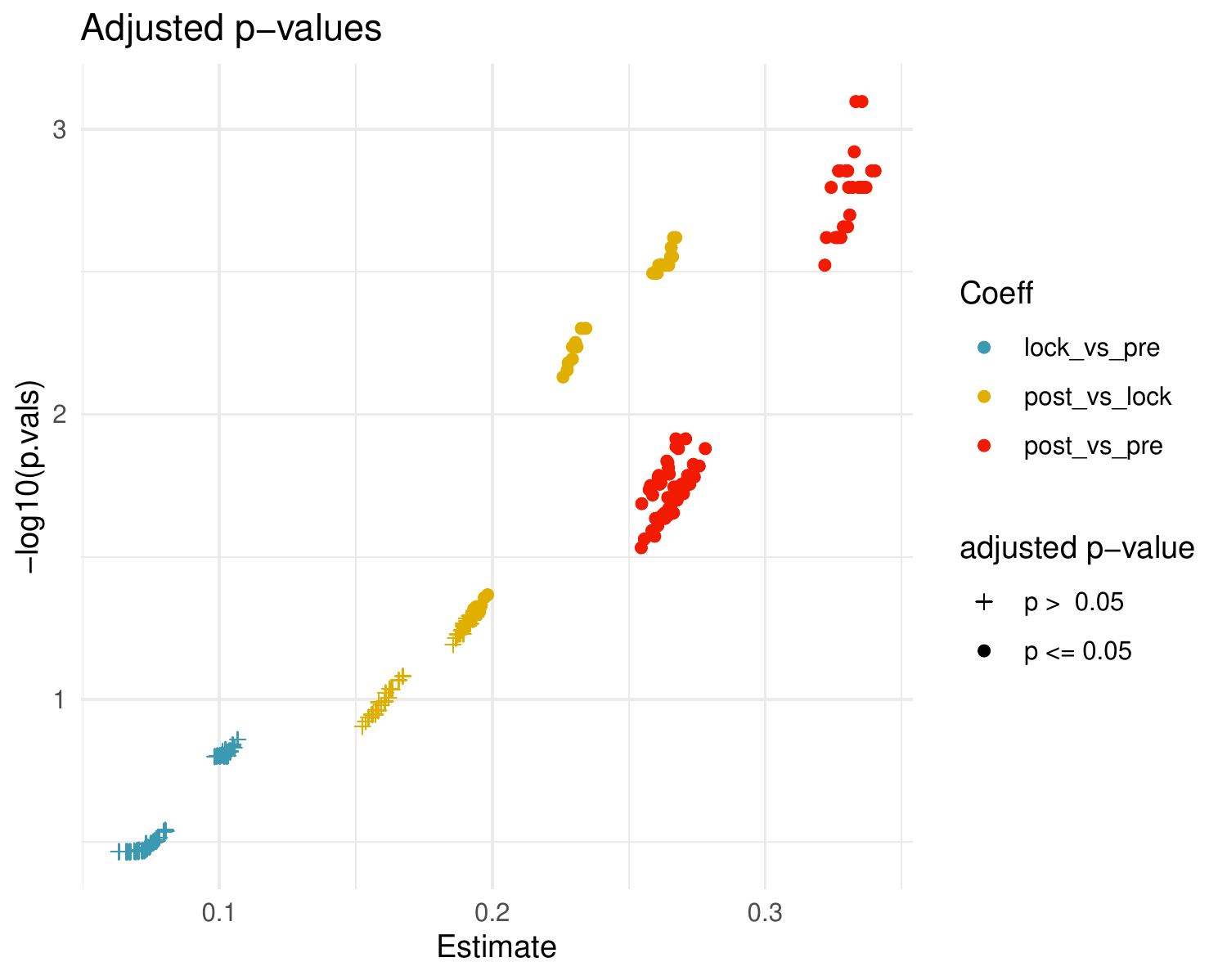}
\caption{
\paolo{Adjusted p-values vs. coefficient estimates of the post-hoc comparisons of 81 tested models in the PIMA of COVID-19 vaccine hesitancy dataset.}}
\label{fig:scatter_adj}
\end{figure}

% --------------------------------------------------
We now move from the descriptive analysis to the inferential one. The global test with post-hoc correction is shown in Table \ref{tab:posthocensemble}. The comparisons {\tt Post} - {\tt Pre} and {\tt Post} - {\tt Lockdown} are significant (overall, over the 81 models of the multiverse), while the comparison {\tt Lockdown} - {\tt Pre} is not. \anna{We point out that the post-hoc correction is based on the three comparisons, and each comparisons is based on the combination of the 81 models. As an example, claiming that the comparison {\tt Post} - {\tt Pre} is significant allows us to account only for the fact that at least one of the 81 models has non-null coefficient for this comparison (and assuming that all models are correctly specified). 
In order to select the most promising models among these we need to move the multiplicity correction to the levels of each individual models. This is done in Figure \ref{fig:scatter_adj} where the adjusted p-values are reported (the table with the detailed results is reported in the supplementary material).}

%\paolo{In conclusion, an overall significant effect of variable of interest period has been observed and at least one specification is statistically significant. We didn't check the formal validity of each model specification (error distribution, outliers, etc...) implying an additional layer of control on the causal effect assessment with PIMA; however, we remember that the significance of an association is only of the elements necessary for the causality assessment \citep{shimonovich2021assessing}}.

\begin{table}[h!]
\centering
\begin{tabular}{llrrrr}
  \hline
 \textbf{Coeff} & \textbf{Stat} & \textbf{nMods} & \textbf{S} & \textbf{Pr($>|z|$)} & \textbf{p.adj (post-hoc)} \\ 
  \hline
Lockdown - Pre-lockdown  & maxT & 81 & 1.71 & 0.1396 & 0.1396 \\ 
Post-lockdown - Lockdown & maxT & 81 & 3.78 & 0.0008 & 0.0008 \\ 
Post-lockdown - Pre-lockdown & maxT & 81 & 4.43 & 0.0002 & 0.0006 \\ 
   \hline
\end{tabular}
\caption{\paolo{Pairwise comparisons of the maxT global test between periods with post-hoc correction in the PIMA of COVID-19 vaccine hesitancy dataset.}}
\label{tab:posthocensemble}
\end{table}

%Figure \ref{fig:scatter_adj} shows the results of the adjusted p-values, while the Appendix reports the detailed results.

Finally, Table \ref{tab:TDP} reports the number of true discoveries and the TDP for each comparison. For the comparison {\tt Post} - {\tt Pre} all models show a significant difference (after multiplicity correction), while {\tt Lockdown} - {\tt Pre} does not show any significant effect. The comparison {\tt Post} - {\tt Lockdown} has an intermediate number of significant comparisons ($29/81=36\%$).

\begin{table}[h!]
\centering
\begin{tabular}{llrrr}
  \hline
 \textbf{Coeff} & \textbf{Stat} & \textbf{nMods} & \textbf{True Discoveries} & \textbf{Proportion} \\ 
  \hline
LockDown - Pre & maxT & 81 & 0 & 0\% \\ 
Post - LockDown & maxT & 81 & 29 & 36\% \\ 
Post - Pre & maxT & 81 & 81 & 100\% \\ 
   \hline
\end{tabular}
\caption{\paolo{Lower $0.95$-confidence bound for the number of true discoveries in each comparison in the PIMA of COVID-19 vaccine hesitancy dataset.}}
\label{tab:TDP}
\end{table}

% ----------------------------------------------------------------------------
% ----------------------------------------------------------------------------

\section{Conclusion and final remarks}\label{sec:conclusion}
In this paper we propose PIMA, a formal inferential framework to multiverse analysis \citep{steegen2016increasing}. Our approach allows researchers to move beyond a descriptive interpretation of the results of a multiverse analysis, and extends other methods to summarize the multitude of performed analyses, such as specification curve \citep{simonsohn2020specification} and vibration analysis \cite{vibrationofeffects} to any generalized linear model. By extending the sign-flip score test  \citep{hemerik2020robust,stdScore} to the multivariate framework, researchers can now make use of the full variety of multivariate and multiple testing methods based on conditional resampling to obtain: 1) weak control of the FWER to test if there is an ‘overall' effect in one of the models explored in the multiverse analysis; 2) strong control of the FWER (i.e., adjusted p-values for each tested model) that allows the researcher to select the models that show a significant effect; 3) a lower confidence bound for the proportion of true discoveries among the tested models. 
Furthermore, PIMA proves to be very robust to over/under-dispersion, allowing for a wide range of models and possible data pre-processing.

This flexibility, however, does not exempt the researcher from the responsibility of the analysis. Some further remarks and considerations in this direction should be made. 

{\bf Define your theoretical model.} In the multiverse analysis framework, each specification should follow a model \paolo{that is based} on a strong theory \paolo{developed within} a research field (e.g. psychology, medicine, physics, etc.). As an example, in \paolo{epidemiology}, a researcher usually defines a set of variables called ‘confounders' in order to adjust the estimated effect between the dependent variable (outcome) and the independent variable (determinant) \paolo{in a} quasi-experimental design. In this case, each specification can be plausible if it includes the same set of confounders as an evaluation of the same initial theoretical model. The exclusion of some confounders is normally used in sensitivity analysis, but it could lead to implausible specifications since the potential mismatch with the theoretical initial model.

{\bf Plan your analysis in advance.} It is important to note that the PIMA method is not iterative, that is, the analysis specifications must be planned before performing the multiverse analysis. Not doing so (i.e. adding or removing models after having seen the results) will add a layer of data manipulation, which is impossible to model and hard to formalize, and therefore can inflate the type I error rate. % A further point is the selection of models to use in the analysis. 
The multiverse approach allows the researcher to plan (in advance) a plethora of models to be explored instead of just one single prespecified model. However, it is still recommended to pre-register PIMA before it is performed.

{\bf Be parsimonious.} There are virtually no limits to the number of models to use, as the proposed PIMA approach will integrate all the resulting information. However, the power will be affected by these choices. Indeed, the overall power to find a significant effect depends on the power of every single model. Although adding 'futile' models will not decrease the quality of false positive control, it will decrease the power of the global test and therefore the ability to detect significant effects. 

{\bf Be exhaustive.} There is a further consideration that applies not only to our inferential methods but also to descriptive methods such as multiverse analysis, specification curve and vibration analysis (and to data modeling, more broadly). When planning the data transformations, the practitioner must realize that failing to take into account any relationship between confounders and the response variable may be a catastrophic source of false positive results.  This case is very well covered in any basic course in statistical modeling, but it may be useful to provide a flavor of the consequences of an inaccurate choice of models in the analysis in practice. We run a simple simulation under the same linear homoscedastic normal framework described in Section \ref{sec:sims}. In this case, however, we do not include the last two confounders in any of the models. The empirical type I error rate exceeds $0.30$ (nominal level $\alpha=0.05$) in all tested models. As a consequence, the combined model exceeds the nominal level by the same amount. The same behavior can be seen in the parametric approach. As practical advice, we recommend including all potential confounders in all models, since losing control of the type I error in any of them will make the inference unreliable. 

A more subtle, but very relevant example is the case where some transformation of the confounders does not account for all the dependence among them. For instance, suppose that a confounder $Z$ has a linear relationship with the response $Y$ and with the variable of interest $X$. Now, to account for non-linear effects, the researcher decides to use a median-split transformation of $Z$. The resulting test on the coefficient of $X$ will lose its control of the type I error. To elucidate this case in practice, we run a second simulation, again under the same setting described above (linear homoscedastic model). In this case, we include all the confounders, but we use a median-split transformation instead of the parabolic models. With sample size $n=200$, the empirical type I error of the true (linear) model is under control (sign-flip score test: $0.051$, parametric: $0.054$), while it largely exceeds $5\%$ for any other model that median-splits the predictors, reaching $0.211$ for the sign-flip score test (and $0.219$ for the parametric test) when the model has all the three confounders with a median-split. As a direct consequence of the loss of control of type I error of the univariate models, the PIMA method loses its error control as well. The empirical type I error is $0.180$ for the maximum, and similar for other combining functions). It would be easy to define more dramatic scenarios, of course. 

%These simple examples stimulates a few further considerations. The significance of each test in a descriptive step (e.g. Multiverse analysis) should be discussed with enormous care: 'Does this low p-value comes from a reasonable model? Are we dealing with dependencies in an appropriate why?'. Therefore, a bag of significant test is not trilling as long as one does not take the responsibility to assume that each one comes from an appropriate model. 
{\bf Thorough discussion of the results.}
The previous consideration may be discomforting. It implies that every significant test must be evaluated with great care and the researcher must take the responsibility to assume that the confounders in each tested model are dealt with properly. However, this is inherently false in many cases. A trivial example comes from the setting of the last simulation above: when a linear relationship is appropriate, the median-split transformation will not provide a test with an adequate control of the type I error. And vice versa, when the dependence should be modeled via a median-split, the natural variables will fail as well. The same can be said for very well known transformations such as log and square-root functions.
These consideration shed a light on the implicit complexity of a multiverse analysis. A significant test must be interpreted as a significant relationship between the predictor of interest $X$ and the response $Y$, given the confounders $Z$ of the model. And the significant result may be due to a real relationship between $X$ and $Y$ or a poor modeling of $Z$. It is the responsibility of the researcher to consider both options carefully.

Let's go back to the application in Section \ref{sec:dataanalysis}. The comparison {\tt LockDown - Pre} does not show any significant result, therefore no (false) claims can be made. More interestingly, the comparison {\tt Post - LockDown} has 29 significant -- i.e. multiplicity corrected -- tests. Let's now focus on this comparison. By exploring the results we can see that most of the significant ones are due to models where the variable {\tt age} is not transformed (i.e. 27/29), while when the age is modeled by a B-spline transformation, the test becomes not significant in most of the cases (see Table \ref{tab:tab_res_age}).
% latex table generated in R 4.1.3 by xtable 1.8-4 package
% Sat Sep 03 20:34:37 2022
\begin{table}[h!]
\centering
\begin{tabular}{lrr}
  \hline
\textbf{Predictor of} {\tt age} & \textbf{p-adjusted} $> 0.05$ & \textbf{p-adjusted} $\leq 0.05$ \\ 
  \hline
Age in continuous &   0 &  27 \\ 
Age with 3 basis of B-splines &  27 &   0 \\ 
Age with 4 basis of B-splines &  25 &   2 \\ 
   \hline
\end{tabular}
\caption{Number of models with significant difference {\tt Post - Lock} for different transformations of the variable {\tt age} in the PIMA of the COVID-19 vaccine hesitancy dataset.}
\label{tab:tab_res_age}
\end{table}
Such a result should cast doubt on the robustness of the results. Most likely, the significant results are due to inadequate modeling of the relationship of {\tt age} with the response variable which, in turn, induces a spurious correlation of the contrast under test with the response variable. In our opinion, therefore, there is not enough evidence to support the claim that the willingness to get vaccinated $Y$ has changed between the  {\tt Pre} lock-down and the {\tt LockDown} period. \anna{This highlights the importance of the multiplicity correction in PIMA to obtain a better understanding of the analysis and results. Particular patterns could suggest, for instance, that significant specifications correspond only to specific modelling choices; the researcher should then evaluate if those particular choices are indeed plausible or, on the contrary, should be discarded.}

{\bf Robust analysis is still possible.}
Despite the challenges pointed out in this discussion, we claim that robust results can still be obtained. Consider the comparison {\tt Post - Pre}, where all comparisons turn out to yield significant effects. If we can assume that ‘at least one' among all tested models deals properly with confounders, we are allowed to claim that there is a significant difference between {\tt Post} and {\tt Pre} -- even though we cannot claim which model is the more appropriate one. This result directly follows from Berger’s general results on intersection-union tests \citep{BergerIntersectionUnion}. Thus, to control the relevant type I error probability it is only necessary to test each one of the coefficients at the $\alpha$ level.
\bigskip

To conclude, we hope our proposed inferential framework for multiverse analysis will allow researchers to learn as much as possible from the multiverse analyses they perform. Our extension to generalized linear models allows researchers who design a greater variety of studies to move beyond a purely descriptive interpretation of a multiverse analysis, and permits researchers to test whether the null hypothesis can be statistically rejected in any or a subset of the models. The strong control for multiplicity in PIMA provides researchers with a statistical tool that allows them to claim that the null hypothesis can be rejected for each specification that shows a significant effect, with the comfort of knowing that they are not p-hacking. \paolo{We underline how the correction for multiple testing is essential, in scenarios with a large number of specifications, to understand for which specifications exactly there is a statistically significant relationship, and so gain a better insight of the analysis.} PIMA makes it possible for researchers who feel that they can not a-priori specify a single analysis approach to efficiently test a plausible set of models while still drawing reliable inferences.

% We hope PIMA increases the usefulness of multiverse analysis for researchers who work in fields where there is uncertainty about the best statistical test of a hypothesis, and who want to transparently and reliably perform a large number of variations of an analysis.}

%%%%%%% old version
%Our proposal of an inferential framework for the multiverse analysis aims to fill the gap between the goals of the multiverse analysis and the lacking of appropriate and comprehensive statistical methodology for doing that. Our extension to generalized linear models is a fundamental step for its generalization and development. In addition, the possibility of choosing a single statistical model can represent a return to the origin, given the control for multiplicity.

\vspace{\fill}\pagebreak

%% ITEM 9 [See the "howto.tex" file.]
%\appendix
%\renewcommand{\theequation}{A\arabic{equation}}
%\setcounter{equation}{0}
%\renewcommand{\thesection}{\Alph{subsection}}
%\setcounter{section}{0}

%\end{table}

%\section*{Appendix A}
%\section*{Appendix B}
%\vspace{\fill}\pagebreak

%% ITEM 10 [See the "howto.tex" file.]
\medskip
\bibliographystyle{apalike}
\bibliography{refs.bib}
%% ITEM 11 [See the "howto.tex" file.]
%%%% You can put your Figures and Tables here
%%%% after the Reference Section.
%%%% BE SURE TO MARK IN THE TEXT WHERE
%%%% YOU WANT EACH FIGURE AND TABLE TO BE PLACED.
%%%% If you prefer, you can integrate your figures and tables into the text of your paper,
%%%% PROVIDED you will provide camera-ready copies of each figure.
%\vspace{\fill}\pagebreak
%\linespacing{1}

%\section*{Figures}
%
%\begin{figure}[h]
%\centerline{\includegraphics{figure01.eps}}
%\caption{Your figure caption goes here.}
%\end{figure}
%\vskip6pt
%\vspace{\fill}\pagebreak

%\section*{Tables}

%\vspace{\fill}\pagebreak
\end{document}